\documentclass[a4paper,fleqn,usenatbib]{mnras}

\usepackage{newtxtext,newtxmath,multirow}
\usepackage{subfigure}
\usepackage[T1]{fontenc}
\usepackage{ae,aecompl}

\usepackage{graphicx}	% Including figure files
\usepackage{amsmath}	% Advanced maths commands
\usepackage{amssymb}	% Extra maths symbols

\def\gtsima{$\; \buildrel > \over \sim \;$}
\def\ltsima{$\; \buildrel < \over \sim \;$}
\def\gsim{\lower.5ex\hbox{\gtsima}}
\def\lsim{\lower.5ex\hbox{\ltsima}}
\title{Ideal Engine Durations For GRB-Jet Launch}

\author[Hamidani, Takahashi, Umeda, \& Okita]
{Hamid Hamidani\thanks{E-mail: hamid@astron.s.u-tokyo.ac.jp}, Koh Takahashi, Hideyuki Umeda, and Shinpei Okita
\\
Department of Astronomy, School of Science, The University of Tokyo, Hongo, Bunkyo-ku, Tokyo 113-0033}
\date{Last updated 25 April 2017}

\pubyear{2017}

\begin{document}
\label{firstpage}
\pagerange{\pageref{firstpage}--\pageref{lastpage}}
\maketitle

\begin{abstract}
Aiming to study GRB engine duration, we present numerical simulations to investigate collapsar jets. We consider typical explosion energy ($10^{52}$ erg) but different engine durations, in the widest domain to date from 0.1 to 100 s. We employ an AMR 2D hydrodynamical code. Our results show that engine duration strongly influences jet nature. We show that the efficiency of launching and collimating relativistic outflow increases with engine duration, until the intermediate engine range where it is the highest, past this point to long engine range, the trend is slightly reversed; we call this point where acceleration and collimation are the highest ``sweet spot'' ($\sim$10 -- 30 s). Moreover, jet energy flux shows that variability is also high in this duration domain. We argue that not all engine durations can produce the collimated, relativistic and variable long GRB-jets. Considering a typical progenitor and engine energy, we conclude that the ideal engine duration to reproduce a long GRB is $\sim$10 -- 30 s, where the launch of relativistic, collimated and variable jets is favored. We note that this duration domain makes a good link with \citet{2013MNRAS.436.1867L}, which suggested that the bulk of BATSE's long GRBs is powered by $\sim$10 -- 20 s collapsar engines.
\end{abstract}


%%%%%%%%%%%%%%%%%%%%%%%%%%%%%%%%%%%%%%%%%%%%%%%%%%

\section{Introduction}

Gamma-Ray Bursts (GRBs) are regarded as the most luminous explosions in the universe \citep{2006RPPh...69.2259M}. Since their discovery \citep*{1973ApJ...182L..85K}, their energy output, temporal properties, cosmological distances and intrinsic properties made them very enigmatic and highly debated. GRB's prompt emission consists of hard photons ($>$ 10 keV) over a very short time scale ($\sim$ 0.1 -- 1000 s). GRBs' duration distribution is bimodal suggesting two-district class of GRBs: Short GRBs (SGRBs; < 2 s), and Long GRBs (LGRBs; > 2 s) \citep{1993ApJ...413L.101K}.

An on-axis observation of a highly relativistic ($\Gamma$ $\sim$ 100) and well-collimated jet (5 -- 10$^{\circ}$) can explain GRBs observational properties, and bypasses the challenging compactness problem \citep{2000PhR...333..529P}. The collapsar model presents a solid explanation to LGRBs, associating them to the death of massive stars, a common phenomenon in the universe \citep*[hereafter MW99]{1993ApJ...405..273W,1999ApJ...524..262M}. The model explains the energy output with a highly rotating and accreting Black Hole (BH). The BH is created as the iron core of rapidly rotating Wolf-Rayet star gravitationally collapses (MW99). Accretion onto the BH produces considerable amount of energy ($\sim$ $10^{52}$ erg; through MHD or neutrino annihilation process), powering two polar jets; jets breakout out from the progenitor and accelerate to reach highly relativistic Lorentz factors (MW99). This model has two essential requirements: i) LGRBs must be associated with high SFR regions in the universe, as they must be related to deaths of massive stars; ii) Metallicity should be low, in order to produce the rapidly rotating BH. Both requirements have been confirmed by observations of GRBs' host galaxies \citep*{2009ApJ...691..182S}, although there have been some rare exceptions \citep[such as GRB020127;][]{2007ApJ...660..504B}. Furthermore, observations of LGRBs confirmed their link to SNe explosions, and thus to deaths of massive stars, for instance: GRB980425/SN1998bw \citep{1998Natur.395..672I} and GRB030329/SN2003dh \citep{2003Natur.423..847H}. Such SNe were categorized into a class of rare and extremely powerful SNe, called Hypernovae (or Broad Line SNe) whose explosion energy is $\sim$ $10^{52}$ ergs \citep[and the references within]{2006NuPhA.777..424N}. Finally, another confirmation to the collapsar model is special-relativistic numerical calculations of massive Wolf-Rayet star models, in particular showing the successful breakout and launch of highly relativistic and collimated jets (\citealt{2000ApJ...531L.119A}; \citealt*{2003ApJ...586..356Z}). Hence, the general popularity and acceptance of the collapsar model in GRB community.

The collapsar scenario for GRBs and their central engines have been investigated in both analytical and numerical approaches. \citet{2003MNRAS.345..575M} analytically investigated constraints on the progenitors in the collapsar model, in terms of GRB duration or driving mechanisms. He argued that He or CO Wolf-Rayet stars are the plausible progenitors, as the duration of many LGRBs necessitates a compact progenitor $\sim$ $10^{10}$ cm. \citet{2011PASJ...63.1243N} carried out general relativistic magneto-hydro-dynamical (MHD) simulations. Studying collimated jets launched from a rotating BH via MHD process, he confirmed that more rapidly rotating progenitors launch more energetic jets. Many similar (and different) studies have since been undertaken, but the detailed properties of the central engine are still far from understood.

On the other hand, a commonly used numerical method consists of considering an inner boundary at which typical engine's relativistic outflow is injected, without including the engine in the computation domain. With this simplified method, GRB jets have been investigated independently from the central engine specific mechanism. Many numerical simulations have been carried-out in this way considering an injection nozzle (e.g. \citealt{2000ApJ...531L.119A}; \citealt{2003ApJ...586..356Z}; \citealt{2005ApJ...633L..17U}; \citealt{2006ApJ...651..960M}; \citealt*{2009ApJ...699.1261M}; \citealt*{2011ApJ...732...26M}; \citealt*{2013ApJ...777..162M}; \citealt*{2007ApJ...665..569M}; \citealt*{2010ApJ...723..267M}; \citealt*{2007RSPTA.365.1129W}; \citealt{2005MNRAS.357..722L}; \citealt*{2007RSPTA.365.1141L}; \citealt*{2009ApJ...700L..47L}; \citealt*{2010ApJ...717..239L}; \citealt{2012ApJ...750...68L}; \citealt{2013ApJ...765..103L}; \citealt{2011ApJ...731...80N}; etc.). These studies investigated the dynamics of a collimated and relativistic jet drilling stellar mantle. This widely used method allows for a comparison of jet properties with basic observational GRB properties. Thus, one may constrain central engine temporal, angular, and energetic properties. 

However, most major 2D hydrodynamic simulations have focused on relatively long engine duration models, in the range of $\sim$10 - 100 s, with most engine duration models $\ge$ 50 s. \citet{2003ApJ...586..356Z} used engines durations > 10 s, and studied the initial parameters for jet propagation inside the progenitor star. \citet{2005ApJ...633L..17U} used a 9 s engine to investigate Hypernova GRB. \citet{2006ApJ...651..960M}, considering 10 s injection duration, studied the effect of the initial $\Gamma$ factor and the initial specific internal energy on the jet properties, such as angular and relativistic properties. He pointed out that transition from GRBs to XRFs (or \textit{low luminosity} GRBs) could be explained by difference in initial specific energies. \citet*{2007ApJ...665..569M} (hereafter ML07), one major study, considered a 50 s injection in order to study temporal and angular properties of the jet. Three phases were found: precursor, shocked and unshocked. ML07 considered the possibility of observing dead times in the GRB light curve, as the shocked phase is narrow and can not be observed at larger viewing angles. \citet{2009ApJ...700L..47L} considered the same jet initial conditions as ML07, including the same 50 s engine duration, to study radiation efficiency using the photospheric model. \citet{2010ApJ...723..267M} considered, again, the same 50 s duration to explain GRBs variability by comparing uniform, variable entropy, and variable baryon load engine models. \citet{2010ApJ...717..239L} used the same engine as well (50 s), to investigate possible SGRB with an off-axis observation of a collapsar jet. Using a 30 s engine, and the photospheric model to derive a thermal prompt emission, \citet{2011ApJ...731...80N} focused on the timing of jet injection in a rapidly rotating massive star, and its effect on the prompt emission. \citet{2011ApJ...732...26M} \& \citet{2011ApJ...732...34L}, both used the photospheric model to derive the GRB emission, for an engine of 100 s. \citet{2013ApJ...765..103L} used different models, most of them with an injection duration of 100 s. Combined with different progenitors and viewing angles, \citet{2013ApJ...765..103L} could successfully populate the same region of the Amati relation \citep{2002A&A...390...81A}. \citet{2013MNRAS.436.1867L} studied the duration of the produced prompt emission in relation to the engine duration, using central engine duration  > 10 s. One of their most interesting results is that BATSE LGRBs engines are on average $\sim$ 20 s long, and that long engines $\sim$100 s must be rare. \citet{2013MNRAS.436.1867L} also argued that even shorter engines are to be considered. Accordingly \citet*{2014MNRAS.442.2202L} considered a uniform 20 s engine and 40 s variable engines, to study the effect of variable engines and how it can justify the observed variability observed in GRB light curves.

Although most of the hydrodynamical simulations focused on long injection duration, only a few short engines models have been studied so far. One rare study is \citet{2009ApJ...699.1261M} where the injection duration was less than 10 s (4 s) to investigate the angular energy distribution of a GRB jet, using different progenitors. Another case is \citet{2012ApJ...750...68L} where injection (from 2 to 15 s) was carried out to study the velocity of the ejecta. Although the central engine and its duration is one very important and not well-understood ingredient in GRB theory, its duration was not widely studied (least of all, in a wide duration domain down to $\sim$ a few s). 

Except in \citet{2012ApJ...750...68L} the question of how engine duration affects the GRBs was not investigated. Furthermore, with the very different engine durations considered so far, there was no study to globally show engine durations in which typical GRBs launch is favored or disfavored. Throughout this paper, we imply with ``typical GRBs'' standard GRBs detected by high-energy instruments (Swift in particular). Such GRBs show a highly variable prompt emission and are located at high redshifts implying extreme energies and luminosities.

Despite decades of progress, many issues related to the mechanism of collapsar engines remain unsolved. This is largely because the physical complexity of the phenomenon. However, neither collapsar model nor the observations \citep{2013MNRAS.436.1867L} seems to exclude engines having more diverse or shorter timescales, at least in some particular conditions of: Rotation, magnetic field, metallicity, and other physical parameters. One strong support of the diversity of collapsar engines in terms of duration is \citet*{2009ApJ...704..354H}. The study shows that the lifetime of the accretion disk (and thus the activity of the engine and its intensity) strongly depends on the rotation of the progenitor, with slowly rotating progenitors powering short intense jets and rapidly rotating progenitors producing long mild jets. Also, the only fact that GRBs are detected over a very wide range of redshift, more than any other known event, implies dramatically different epochs and environments. Accordingly, theoretical timescale of collapsar engines must be wide to reflect such diversity and natural differences. Furthermore, one notable study on collapsar engine durations, \citet{2012ApJ...749..110B} gives some insights on the distribution of GRB engines. \citet{2012ApJ...749..110B} found that the duration distribution at long durations ($\sim$100s) can be fitted with a power law index --4 < $\alpha$ < --3; thus suggesting that longer engines must be rare (for more information about the methodology see: \citealt*{2011ApJ...739L..55B} and \citealt{2011ApJ...740..100B}). \citet{2012ApJ...749..110B} also pointed that this power law, if extrapolated down to shorter durations, predicts that there are more intermediate than long engines, and much more short engines than intermediate and long engines; thus many chocked than successful jets, and so, more \textit{low luminosity} GRBs relative to regular GRBs, which would explain the huge rates of \textit{low luminosity} GRBs relative to regular GRBs as suggested by observations (see: \citealt{2005MNRAS.360L..77C}; \citealt{2006Natur.442.1011P}; \citealt{2006Natur.442.1014S}; \citealt{2006ApJ...645L.113C}; \citealt{2007ApJ...662.1111L}; \citealt*{2007ApJ...657L..73G}; etc.) Considering this possibility (although there is no evidence) and the actual power law at longer engines, we constructed our engine duration domain. Thus, we study collapsar jets with shorter engine duration along with intermediate and long durations, in a search for engine durations where GRBs are favored.

In this study, we consider a series of numerical simulations, using a very wide variety of engine durations (0.1 -- 100 s) extending ML07 study of jet's hydrodynamic phases, and exploring some engine durations that have not been numerically investigated yet. For our different engines we consider the same total injected energy, $10^{52}$ erg. In other words, we consider the engine duration as a parameter backed by diversity in nature. The choice of total energy is in accordance with observations of Hypernovae explosions \citep{1998Natur.395..672I, 2006NuPhA.777..424N}, as well as typical for hyperaccretion as in collapsar theory (MW99). 

For this, we use an Adaptive Mesh Refinement (AMR) two-dimensional special relativistic hydrodynamical numerical code. We use deeper injection nozzle position, $10^8$ cm instead of the $10^9$ cm used in ML07 and most previous studies. An injection nozzle of $10^9$ cm, might simplify the jet interaction with the star, while $10^8$ cm is more realistic, bringing the star-jet interaction to a higher and more realistic degree (although such a calculation is about one order of magnitude more time-consuming). We investigate how jet phases, precursor, shocked and unshocked, and other temporal, angular and energetic properties, depend on the injection duration. We finally discuss the nature of the observed GRBs for our different engine models. The numerical simulations were carried-out using CFCA X30. In total, this study consumed about 600 000 core-hours of computing time.

This paper is organized as follows: In $\S$ \ref{sec:Numerical method}, we explain the method and formalism used to perform the simulations. In $\S$ \ref{sec:Setup of the simulations}, we explain the stellar model, the grid system and the jet initial conditions for each model. In $\S$ \ref{sec:Data processing procedure}, we explain the method used to derive the angular and energetic properties. Results are presented and discussed in $\S$ \ref{sec:Results}. A conclusion is given at the end of this paper ($\S$ \ref{sec:Conclusion}).

\section{Numerical method}
\label{sec:Numerical method}

We performed numerical simulations with a two-dimensional relativistic hydrodynamical code, the same code used in \citet*{2014MNRAS.438.3119Y} and \citet{2012AIPC.1484..418O}, with a newly added AMR treatment as described in $\S$ \ref{subsec:Grid}. We assume the explosion and jet propagation to be axisymmetric. The basic equations solved in this code are given as:

\begin{eqnarray}
    \frac{\upartial U}{\upartial t} + 
    \frac{1}{r^2} 
    \frac{\upartial (r^2F^r)}{\upartial r} + 
    \frac{1}{r \sin \theta} 
    \frac{\upartial(\sin \theta F^{\theta})}{\upartial \theta} = S + G
\end{eqnarray}

Where $U, F^i, S$ and $G$ are conserved vector, i-component of numerical flux, source term and gravitational source, respectively. Under geometrical unit $G = c = 1$, where $G$ and $c$ are the gravitational constant and the speed of the light, these vectors are written as follows \citep[e.g.][]{2005A&A...436..503L}:

\begin{eqnarray}
U = (\rho \Gamma,\rho h \Gamma \upsilon^r,\rho h \Gamma \upsilon^{\theta},\rho h \Gamma^2 - p - \rho \Gamma)
\end{eqnarray}

\begin{equation}
F^i=(\rho \Gamma \upsilon^i,\rho h \Gamma^2 \upsilon^{i+r}+\delta_{\rm r}^{\rm i},\rho h \Gamma^2 \upsilon^{i+\theta}+p\delta_{\rm \theta}^{\rm i},\rho h\Gamma^2 \upsilon^i - \rho \Gamma \upsilon^i)
\end{equation}

\begin{eqnarray}
S = \tfrac{1}{r} (0,\rho h \Gamma^2 \upsilon^{\theta} \upsilon^{\theta} + p, - \rho h \Gamma^2 \upsilon^r \upsilon^{\theta}, 0)
\end{eqnarray}

\begin{eqnarray}
G = (0, \rho h \Gamma \upartial_r \Phi, 0, \rho h \Gamma \upartial_r \Phi)
\end{eqnarray}

Here, $\rho$, $\upsilon$ and $p$ are the rest mass density, velocity and pressure. $h$ and $\Gamma$ are specific enthalpy and Lorentz factor, respectively, defined as $h=1+\varepsilon+p/{\rho}$, where $\varepsilon$ is specific energy density, and $\Gamma =1/\sqrt{1-\upsilon^2}$. Gravitational potential $\Phi$ includes the contributions of self-gravity and the central remnant. The integration form of the Poisson equation approximated in Newtonian mechanics (Hachisu 1986) is applied for self-gravity. Time integration is calculated using the second-order Runge-Kutta Method developed by \citet*{1988JCoPh..77..439S}. We use a simple equation of state (EOS), the so-called gamma law EOS $p=(\Gamma-1) {\rho} {\varepsilon}$, with an adiabatic index $\Gamma = 4/3$, which accounts for both the gas and radiation components. Our choice of a simple EOS is due to our focus on the jet general properties, which our simple EOS would not overlook.

As long as only penetration through the stellar mantle and propagation in the CSM is considered, the incoming jet can be characterized by several parameters, regardless of the detailed mechanism of central engine. Here, we follow the method proposed in \citet{2009ApJ...690..526T} to determine the boundary condition of ($\rho_0,\upsilon_{r0},p_0$), where the index ``0'' indicates that the quantity is calculated at the inner boundary of the computational domain. Thus, characterizing a jet comes down to defining the following 6 parameters: $T_{inj}, E_{tot}, R_{in}, \theta_{op}, f_{th}$ and $\Gamma_0$. The key parameter of this study is $T_{inj}$, the duration of the energy injection in the simulation, which reflects the engine duration. $E_{tot}$ is the total energy injected, up to $time = T_{inj}$, in other words, it is the total energy released by the central engine in the form of two relativistic polar jets (assumed as $10^{52}$ erg). Thus, the energy deposition rate, assumed constant, can be written as $\dot{E}=E_{tot}/T_{inj}$. $R_{in}$ and $\theta_{op}$ are parameters to determine the geometric property of outflow: the inner boundary where the injecting nozzle is placed and the opening angle of the jet cone, respectively. With these two parameters, we can get the intersection area of the inner boundary and jet cone as $A_0=4 \pi R_{in}^2 (1-\cos \theta_{op})$. Finally, $f_{th}$ is the ratio of the thermal to total injected energy, and $\Gamma_0$ is the Lorentz factor of the outflow at the inner boundary. Once these parameters are set boundary conditions will obtained from the following equations:

\begin{eqnarray}
\upsilon_{r0} = \sqrt{1 - 1/{\Gamma_0^2}} 
\end{eqnarray}

\begin{eqnarray}
p_{0} = \frac{f_{th}\dot{E}}{\upsilon_{r0}A_0(\tfrac{\gamma}{\gamma -1}\Gamma_0^2-1+f_{th})}
\end{eqnarray}

\begin{eqnarray}
\rho_{0} = \frac{ (1-f_{th}) (\tfrac{\dot{E}}{\upsilon_{r0}A_0}-p_0)}{\Gamma_0(\Gamma_0 - 1)}
\label{eq:den}
\end{eqnarray}

As mentioned above, we focus on the properties of GRB central engine with various duration times. As \citet{2013MNRAS.436.1867L} has suggested central engines, on average, are active over several tens of seconds; shorter engines are likely, but very long engines might be rare. Thus, we set engine durations $T_{inj}$ from 0.1 to 100 s, covering such an unprecedentedly wide range. We note that although our models vary in $T_{inj}, E_{tot}$ is always the same. Thus, the input energy (or central engine power: $\dot{E}={E_{tot}}/{T_{inj}}$) also varies according to $T_{inj}$. As the density of the jet material is proportional to $\dot{E}$ (equation \ref{eq:den}), long $T_{inj}$ favors low-density jets and short $T_{inj}$ gives denser jets (for more details see: figure~\ref{fig:snapshots} \& $\S$ \ref{subsec:Phases and breakout properties}). 

In the context of the collapsar scenario, energy conversion efficiency $\eta$ is defined as the ratio of the energy powering the jet, or the input energy, to the rest mass energy accretion rate onto the BH. As we assume these parameters not to vary in time, using an accretion rate $\dot{M}, \eta$ is defined as:

\begin{eqnarray}
\eta = \frac{\dot{E}}{\dot{M}c^2}
\label{eq:acc}
\end{eqnarray}

Since both the time derivative values are considered to be constant in time, total accretion mass $M_{acc}$ can be derived with time integration of equation (\ref{eq:acc}) as:

\begin{eqnarray}
\eta = \frac{E_{tot}}{\eta c^2}
\end{eqnarray}

The right side is only dependent on $E_{tot}$ and independent of $T_{inj}$. Therefore, the assumption of taking a constant $E_{tot}$ and various $T_{inj}$ is equivalent to considering the same total accreted mass but with different accretion rates. Such different accretion rates can be justified by parameters related to the progenitor or the environment (e.g: rotation and metallicity).

\section{Setup of the simulations}
\label{sec:Setup of the simulations}
\subsection{Stellar model}
\label{subsec:Stellar model}

\begin{figure}
 \includegraphics[width=\columnwidth]{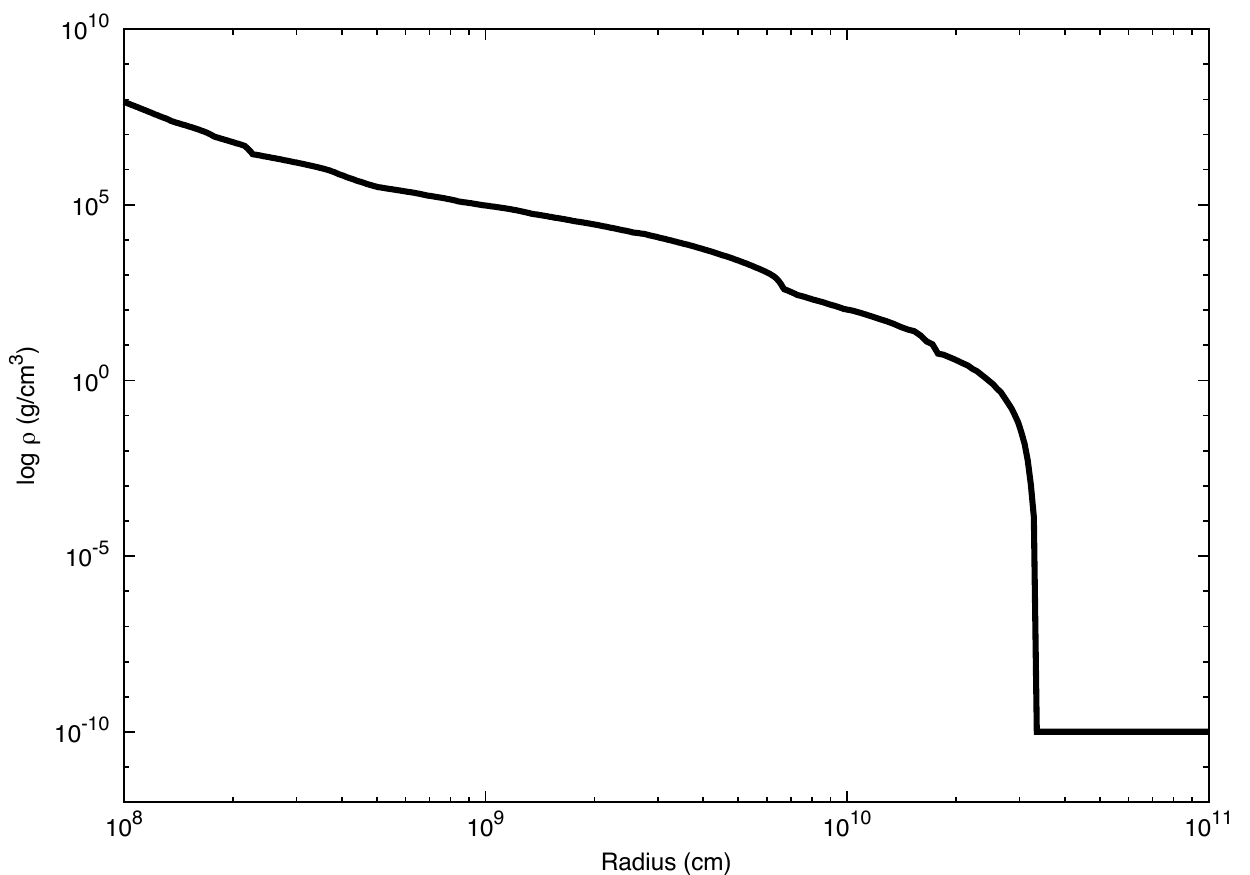}
 \caption{Density profile of the progenitor.}
 \label{fig:progenitor}
\end{figure}

The progenitor model for the simulations presented here is made from a $25M_{\sun}$ initial mass star model as in \citet*{2008ApJ...673.1014U}. The star loses a fraction of its H envelope by mass-loss wind, down to $20.4M_{\sun}$ at the pre-SN phase. We artificially remove the H envelope to make a $6.1M_{\sun}$ of He Wolf-Rayet star for our GRB progenitor, with a radius of $3.3\times10^{10}$ cm (see the progenitor density profile in figure~\ref{fig:progenitor}). Initial pressure and density are taken from the progenitor model. \citet{2003MNRAS.345..575M} argued that such a compact progenitor is preferable, as some of the observed GRBs have short durations, about a few seconds and such short durations must be explained by a compact progenitor. Our stellar model has a relatively low mass, in particular in comparison with the widely used 16TI model of \citet*{2006ApJ...637..914W}. It is on the lower end for a BH forming SN. This choice is due to our intention to focus on moderate events and avoid a bias to particularly extreme GRB progenitors. Relative to 16TI, jets with this model need slightly shorter times and lower luminosities to breakout, which gives a wider spectrum of observable jets (though the orders remain very comparable and the trends are unaffected).

The surrounding medium is taken uniform with a density $\rho=10^{-10}$ g cm$^{-3}$. A Courant Number (CFL) of 0.3 is used for the simulations presented here. The effect of rotation is ignored, since its dynamical timescale is much longer than the timescale of our simulations ($\sim100$ s). Thus we consider that it is safe to assume that our progenitor remained almost static during the simulation. Neutrino pressure and general relativistic effects from the central black hole are not considered either, as the inner boundary at $10^8$ cm, is at about $10^3$ gravitational radii away from the region dominated by this effects, thus both effects could be safely ignored. 

\subsection{Grid}
\label{subsec:Grid}

We use a 2D spherical coordinate system $(r, \theta)$ with axisymmetry and equatorial plane symmetry. Apart from the injection region, the boundary conditions at the polar axis and equatorial plane of the grid are reflective, as we consider that a symmetric jet is emerging in the opposite direction. Computational domain extends from $R_{in} = 10^8$ to $R_{out} = 3.01\times10^{12}$ cm, allowing the relativistic outflow to be followed for about 100 s (for: $0^{\circ} \geq \theta \geq 90^{\circ}$). Radial grid is set to vary in an AMR mode, initially from 2500 up to over some 10 000 for long injection models, following and attributing higher resolutions for the relativistic outflow ($\Gamma$ $>$ 1) accordingly. We use 11 levels of refinement, and the corresponding resolutions varies as:  $\Delta r_l = \Delta r_0\times2^l$, where $l$ corresponds to the level of mesh refinement (from 0 to 10) and $\Delta r_0$ is the lowest resolution ($\Delta r_0 = 10^{10}$ cm, level 0 of refinement). $\Delta r_0$ is considered in the exterior of the star, at regions where there is no relativistic outflow (where $\Gamma$ = 1). The highest radial resolution ($\Delta r_{10} = 9.7\times10^6$ cm, level 10 of refinement) is adopted at the stellar surface, in order not to miss any potential precursor structure. Inside the progenitor, in the innermost region we consider significantly high resolutions ($10^7$ cm $\geq \Delta r \geq 3.9\times10^7$ cm, always higher than level 8). Also, we always follow the jet head in high resolution ($\Delta r = 3.9\times10^7$ cm, 8 levels of refinement). Although our radial resolution is still lower than that of ML07 ($7.8125\times10^6$ cm), it is reasonably good considering most previous studies' resolutions (e.g. \citealt{2011ApJ...732...26M} with $\Delta r_{0} = 10^7$ cm and \citealt{2011ApJ...731...80N} with $\Delta r_{0} = 10^8$ cm).

For the polar grid $\theta$ we employ $N_{\theta} = 256$ uniform logarithmic grids, with angular resolutions varying from $\Delta\theta_0 = 0.088^{\circ}$, at the jet on-axis, to $\Delta\theta_{255} = 0.896^{\circ}$ at the equator, such that: $\Delta\theta_n =  \Delta\theta_0\times C^n$, where $C = 1.009$. This angular resolution is reasonably high in comparison to many previous works (for example $0.25^{\circ}$, for \citealt{2011ApJ...732...26M} and \citealt{2003ApJ...586..356Z}) although the resolution in ML07 is still higher ($0.0358^{\circ}$). 

Our inner boundary is placed at a relatively deeper region, $10^8$ cm, in comparison to most previous studies, that generally used $10^9$ cm \citep[ML07][]{2010ApJ...723..267M, 2009ApJ...700L..47L, 2011ApJ...732...34L, 2013ApJ...765..103L, 2006ApJ...651..960M, 2009ApJ...699.1261M, 2011ApJ...731...80N, 2014MNRAS.442.2202L}. This deep inner boundary is to better capture the evolution of the jet inside the star, especially for the short engines that our study includes. It is also more realistic as it is closer to the region where the central engine is expected to inject energy, near the BH horizon at $\sim10^{6-7}$ cm. Although, simulations with such deep injection consume a lot of computational power, we believe that we still could afford very good resolutions compared to previous studies that used a similar deep injection but at the cost of increasingly poor resolution at large radii, and limited computation domain (e.g. \citealt{2003ApJ...586..356Z} injecting at $2\times10^8$ cm \& \citealt{2000ApJ...531L.119A} at $2\times10^7$ cm).

\subsection{Total energy}
\label{subsec:Total energy}

We assume the collapsar total energy, delivered by the engine in the timescale $T_{inj}$, as $10^{52}$ erg (independently of $T_{inj}$). The reasons of this consideration are as follow: 1) $\sim10^{52}$ erg is reasonably justified by theoretical models of a collapsar engine of 2-3 solar masses (MW99); more than $10^{52}$ erg is possible but with optimistic models or in extreme conditions (e.g. of angular momentum). 2) The best GRB-SN connections so far invoke SNe of Hypernova type, with the explosion energy estimated in the order of $10^{52}$ erg (e.g. 980425/1998bw, 030329/2003dh, 031203/2003lw \& 100316D/2010bh; see \citealt{1998Natur.395..672I} or \citealt{2006NuPhA.777..424N} for a summary); these SNe provide an estimation of the original collapsar explosion energy budget, in the order of $10^{52}$ erg. 3) From observations, GRBs with redshift measurement show isotropic equivalent energy in the range of $10^{51}$ -- $10^{54}$ ergs \citep*{2009A&A...508..173A}; the beaming factor reduce it to the true radiated gamma energy $E_{\gamma}$ by a factor of 1-2 orders of magnitude \citep{2001ApJ...562L..55F}; while typical radiative efficiencies estimated in the range of $\sim$ 10\% to 90\% (see: \citealt{2013MNRAS.428..729M} \& \citealt{2007ApJ...655..989Z}) bring estimations on the energy delivered by engines to at least $\sim$ a few to 10 times $E_{\gamma}$. At the end, an engine energy of $\sim10^{52}$ erg seems very reasonable.

\subsection{Jet conditions and engine models}
\label{subsec:Jet conditions and engine models}

Jet energy is inputted at a radius of $R_{in} = 10^8$ cm from the center of the progenitor, at the inner boundary of our computational domain. The jet initial opening angle is adopted as $10^{\circ}$, as in major previous studies  \citep[ML07,][etc.]{2006ApJ...651..960M, 2013ApJ...765..103L}. According to \citet{2006ApJ...651..960M} who investigated the preferable conditions for GRB jet, hot and mildly relativistic initial jet is required for successfully launching the highly collimated and ultra-relativistic jet necessary to produce GRB. Thus, thermal energy fraction and initial Lorentz factor of the injected jet are taken as $f_{th}=0.975$ (hot) and $\Gamma_0=5$ (mildly relativistic). The maximum Lorentz factor, defined as the terminal Lorentz factor at infinity when all internal energy will be converted to kinetic energy, is $\Gamma_{\infty}\sim160$ (according to Bernoulli relativistic equation: $\Gamma_{\infty}\sim h \Gamma$).

The injection duration is considered from 0.1 to 100 s. To easily analyze the different engine models, we separate them into four groups, from the shortest: ``brief'' engines ($T_{inj} < T_{breakout} \sim$ 2 s; or ``chocked'' jets), ``short'' engines ($\sim$ several seconds; 2 s $\geq T_{inj} < 10$ s), ``intermediate'' engines ($\sim$ several ten seconds; 10 s $\geq T_{inj} < 40$ s) and finally ``long'' engines (50 s $\geq T_{inj} \geq 100$ s). This engine notation will be followed all along this paper. 

\begin{figure}
 \includegraphics[width=\columnwidth]{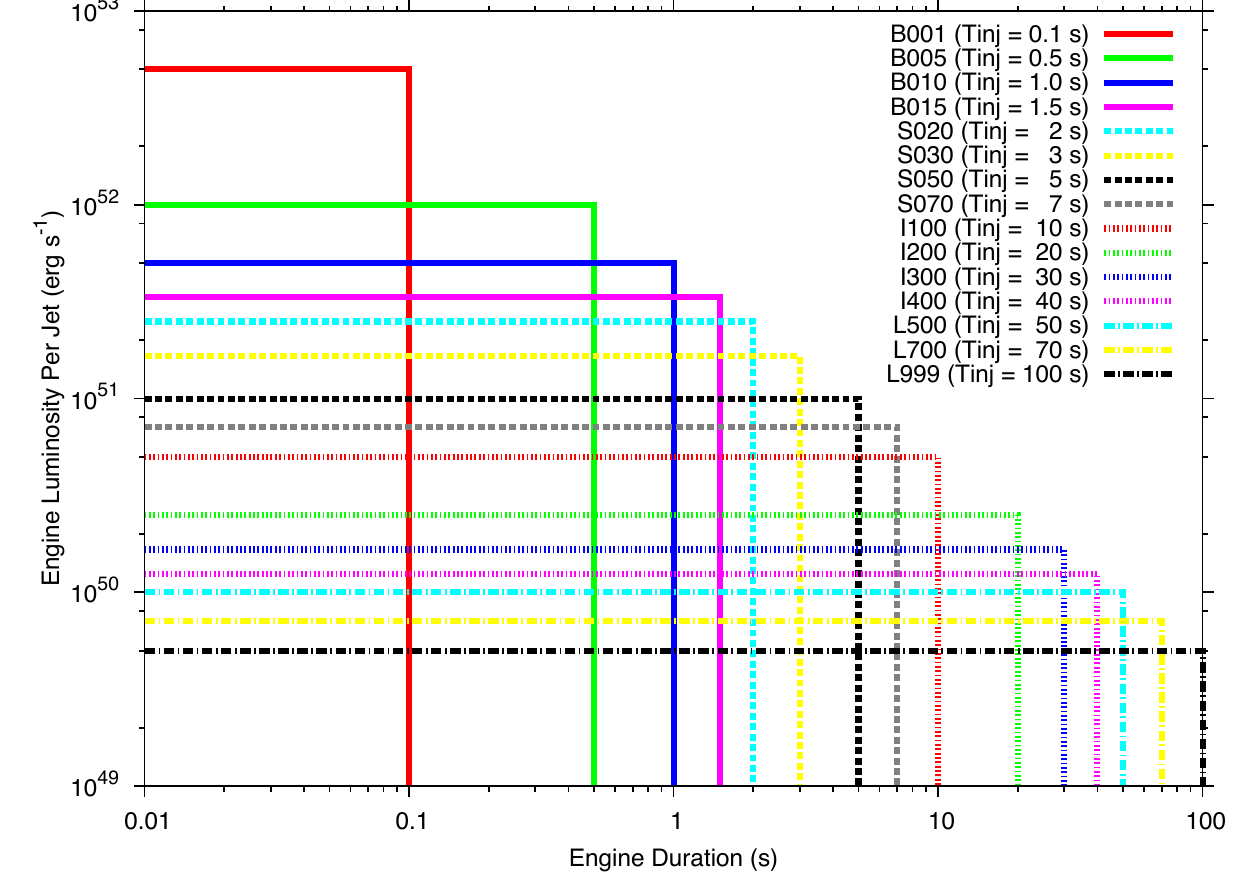}
 \caption{Engine durations and the corresponding luminosities, per jet.}
 \label{fig:engines}
\end{figure}

The computed models are summarized in table \ref{table:engines}, with the corresponding inputted luminosities per jet (Figure~\ref{fig:engines}). As the duration varies over 3 orders of magnitude while the total energy is constant at $10^{52}$ erg, the luminosity of the inputted jet is also very diverse (from $5\times10^{52}$ to $5\times10^{49}$ erg s$^{-1}$), covering intense-short to long-mild engines. This diversity of jet powers reflects the expected natural diversity in progenitors' physical parameters (in particular parameters such as angular momentum; \citealt{2006ApJ...651..960M}).

%%%%

\begin{table}
 \caption{Parameters of the computed models.}
 \label{table:engines}
 \begin{tabular}{llll}
  \hline
  Model & Injection & BH Luminosity & Engine Type\\
 & Time (s) & Per Jet (erg s$^{-1}$) & Type\\
  \hline
B001 & 0.1& $5.0\times10^{52}$ & Brief\\[1pt]
B005	 & 0.5 & $1.0\times10^{52}$ & Brief\\[1pt]
B010	 & 1.0 & $5.0\times10^{51}$ & Brief\\[1pt]
B015	 & 1.5 & $3.3\times10^{51}$ & Brief\\[1pt]
S020	 & 2.0 & $2.5\times10^{51}$ & Short\\[1pt]
S030	 & 3.0 & $1.6\times10^{51}$ & Short\\[1pt]
S050	 & 5.0 & $1.0\times10^{51}$ & Short\\[1pt]
S070	 & 7.0 & $7.1\times10^{50}$ & Short\\[1pt]
I100 & 10.0 & $5.0\times10^{50}$ & Intermediate\\[1pt]
I200 & 20.0 & $2.5\times10^{50}$ & Intermediate\\[1pt]
I300 & 30.0 & $1.6\times10^{50}$ & Intermediate\\[1pt]
I400 & 40.0 & $1.2\times10^{50}$ & Intermediate\\[1pt]
L500 & 50.0 & $1\times10^{50}$ & Long\\[1pt]
L700 & 70.0 & $7.1\times10^{49}$ & Long\\[1pt]
L999 & 100.0 & $5\times10^{49}$ & Long\\[1pt]
16TIg5$^*$ & 50.0 & $5.32\times10^{50}$ & - \\[1pt]
  \hline
 \end{tabular}
 \vspace{1ex}
 
     \raggedright {\footnotesize $^*$ Same parameters as 16TIg5 model in ML07; the simulation was carried out assuming an outer injection nozzle at $10^9$ cm, and an engine total energy of $5.32\times10^{52}$ ergs, while for all the other calculations the injection is at $10^8$ cm, and the engine energy is $10^{52}$ ergs.}
\end{table}

%%%%%%%%%%%%%%%%%%%%%%%%%%%%%%%
%%%%%%%%%%%%%%%%%%%%%%%%%%%%%%%
\section{Data processing procedure}
\label{sec:Data processing procedure}

In this section we explain the method and approximations used for the treatment of numerical data. We adopt the same method as in ML07 (see ML07 $\S$  4.1) to derive jet light curves and angular properties of our relativistic jets. A snapshot of the simulation data is saved every 1/10th seconds of simulation time (1/15th seconds for ML07). As in ML07, the energy flux is determined as a function of angle and time by finding all the points that will cross a given fixed radius within the next 0.1 s. We use the same approximation for sideways expansion as in ML07, by spreading every point's energy equally over an angle of $\pm1/\Gamma_\infty$ from the direction of motion of the fluid at that point. As argued in ML07, this accounts for hydrodynamic spreading and the relativistic beaming of the emitted radiation (ML07). The energy is then placed into the same system of angular bins as in ML07, where the total energy in each angular bin is calculated considering contributions from points at different angles. Finally, only outflow energy above a specified minimum Lorentz factor is considered, excluding any fluid energy with a lower Lorentz factor. In this way, the simulation data from each snapshot file can be added over time, to estimate the total energy seen at a fixed radius from different angles. We consider the same minimum Lorentz factor and radius in ML07, $\Gamma_{min} = 10$, to derive each model's light curve (see ML07's Fig. 12). Both the light curve and the energy angular distribution are calculated for the same minimum Lorentz factor, and at the same radius ($R = 1.2\times10^{11}$ cm). We use 45 angular bins, identical to those considered in ML07, consisting of (from small to large angles, with $0^{\circ}$ at the on-axis region of the jet): 14 bins with an angular width of $0.25^{\circ}$ (with centers ranging from $0.125^{\circ}$ to $3.375^{\circ}$), 17 bins with a width of $1.0^{\circ}$ (from $4.0^{\circ}$ to $20.0^{\circ}$), and finally 14 bins spaced every $5.0^{\circ}$ ($23.0^{\circ}$ to $88.0^{\circ}$).

%%%%%

\section{Results}
\label{sec:Results}

\subsection{Phases and breakout properties}
\label{subsec:Phases and breakout properties}

Previous studies have showed that jet evolution reveals different hydrodynamical phases \citep[ML07, etc.]{2000ApJ...531L.119A, 2004ApJ...608..365Z, 2007RSPTA.365.1141L}. In the following sub-sections we explain the properties of these phases as found in previous studies. Then we present the results of our simulations and discuss phases, their breakout and contribution to jet nature, as a function of the engine duration. 

\subsubsection{The confined phase}
\label{subsubsec:The confined phase}

\begin{figure}
 \includegraphics[width=\columnwidth]{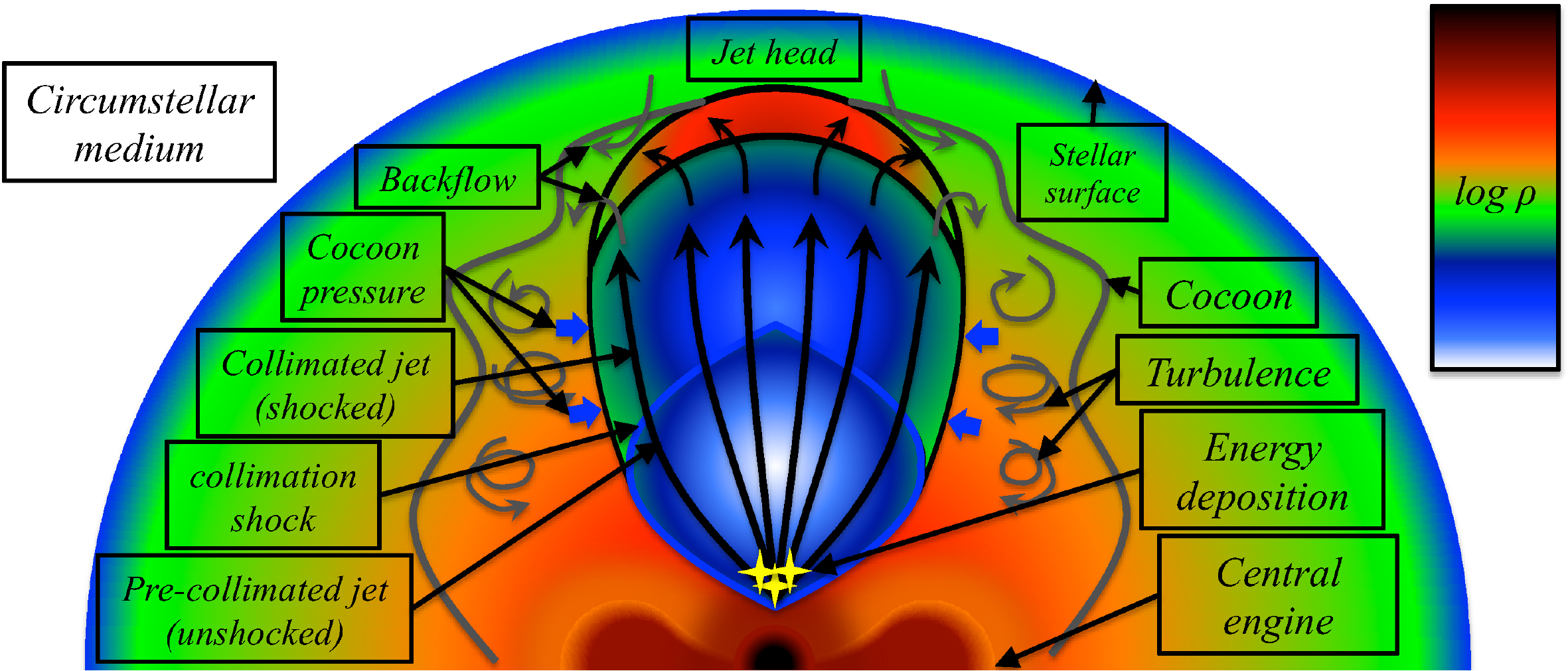}
 \caption{Illustration of a collapsar jet in the confined phase.}
 \label{fig:illustration}
\end{figure}

First, we present a short review the nature of the first phase, which is called ``the confined phase'' \citep{2007RSPTA.365.1141L}. During this phase the jet is confined inside the progenitor, and its head progresses from the injection nozzle to the stellar surface. A high-pressure cocoon is formed behind the jet head, keeping the jet collimated (see figure~\ref{fig:illustration} for an illustration). This phase has no observational signature, and thus it's a non-radiative phase \citep{2007RSPTA.365.1141L}. However, as the premature jet is formed and shaped during this phase, the hydrodynamical properties of the mature jet, developed later, will trace those of this confined jet. Although it has been proved that the speed of the jet head is fairly independent of stellar properties, the energy stored in the cocoon was found as proportional to engine luminosity \citep[see][ML07]{2007RSPTA.365.1141L}. Thus, we expect our different engine models with their different luminosities to produce dramatically different jet structures.
 
\begin{figure}
 \includegraphics[width=\columnwidth]{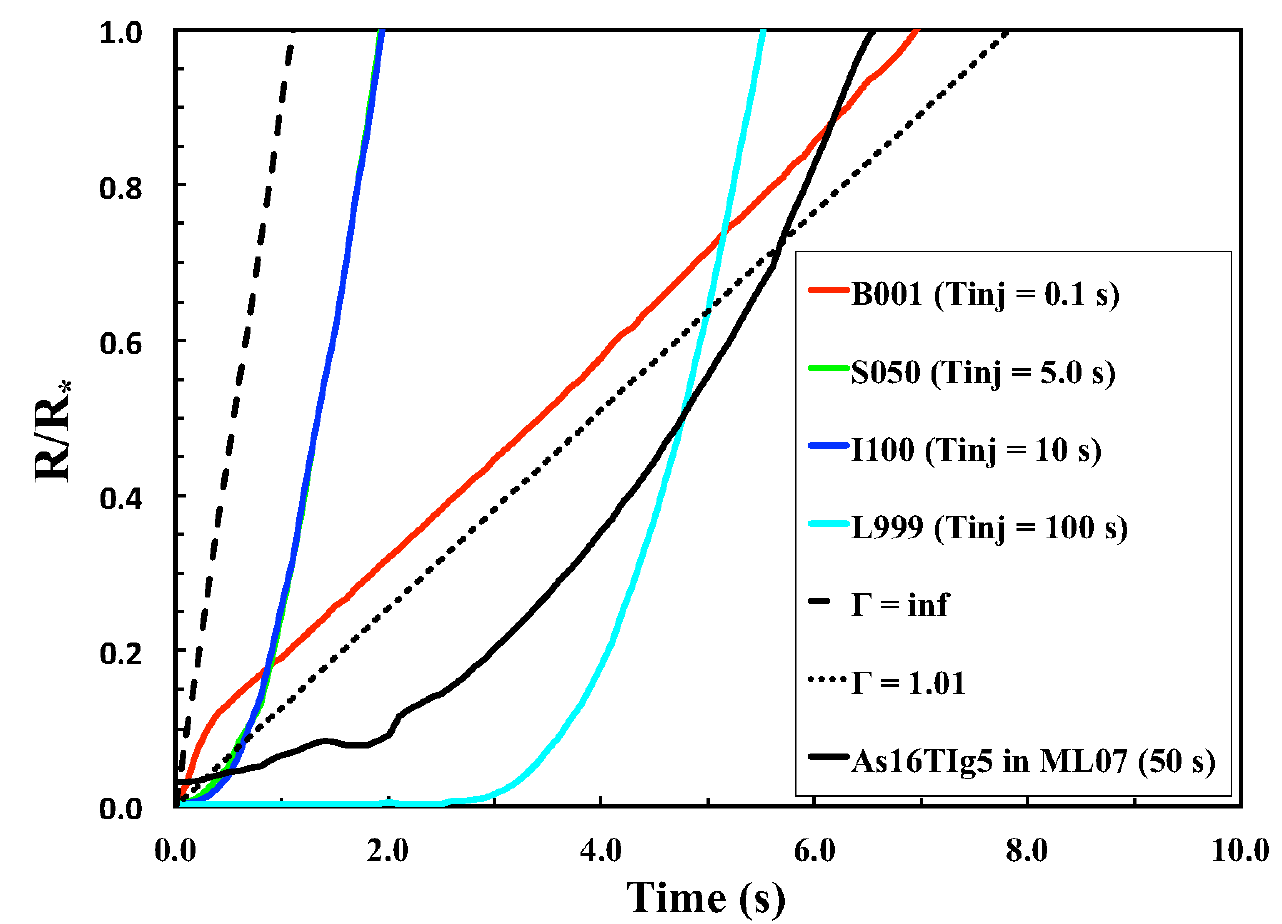}
 \caption{Jet head propagation along the on-axis inside the progenitor before the breakout, for the briefest computed engine model (B001), a short engine model (S050), an intermediate engine model (I100) and the longest engine model considered (L999). An engine model similar to 16TIg5 used in ML07 is also shown. For a comparison, dashed lines shows the maximum allowed speed (c), dotted lines shows the slope of a sub-relativistic speed of a Lorentz factor 1.01.}
 \label{fig:confined jet}
\end{figure}

We present results from our simulations. In figure~\ref{fig:confined jet} we show jet head propagation for some of our engine models, from the briefest B001, to the longest L999, with short and intermediate engine models (S050 and I100). A model similar to that of ML07 (16TIg5), with the same initial Lorentz factor, opening angle, total energy ($5.32\times10^{52}$ erg), and nozzle position ($10^9$ cm) is also shown. In the first moments, and at small radii ($< \sim10\%$ of the stellar radius), brief engine's jet head is ahead expanding at a relativistic speed (with a slope close to that of the speed of light c). While in both short and intermediate models, speed is initially sub-relativistic (with slopes similar to that of a $\Gamma$ = 1.01). Soon after, at larger radii, the brief engine is off, and behaviors are inversed, with jet head in S050 and I100 gradually gaining speed (converging to c), whereas in B001 jet head looses speed and finishes with a roughly constant sub-relativistic speed, until it breaks out. As a consequence jets in short and intermediate models have the shortest breakout times, $\sim2$ s, and the highest Lorentz factors at the moment of the breakout, whereas jets in brief engine models have the longest breakout times, up to 7.0 s, and some of the lowest Lorentz factors (see table \ref{table:breakouts} \& \S \ref{subsubsec:Breakout times}). In L999, jet head evolution is similar to that of S050 and I100, although it takes the jet about $\sim3$ s to breakthrough the inner region. This delay is explained by the significantly lower energy deposition of this engine ($5\times10^{49}$ erg s$^{-1}$). However, later on, once the jet head is effectively launched, it shows an evolution similar to that in S050 and I100, with initially a sub-relativistic speed (or slope) in the inner radii, and then gradually increasing speed at larger radii until the jet head breaks out relativistic. Affected by the initial delay, the jet in L999 model shows a breakout time of 5.6 s, significantly longer than that of jets in S050 and I100. Thus, a comparison based on the engine duration shows that engine duration has a significant influence on the jet head propagation inside the progenitor and its breakout time, mostly with the behavior of brief engines' jets ($T_{inj} < T_{breakout}$) contrasting with that of longer engines (for more details see \S \ref{subsubsec:Breakout times}).

Our replication of ML07's 16TIg5 model \citep[and the uniform model in][]{2010ApJ...723..267M} gave very similar evolution. Although the calculations are still not fully identical, with some differences in the ratio of internal over rest mass energy, progenitor, EOS, resolution, etc. Our 16TIg5 simulation has a breakout time of 6.8 s, in ML07's 16TIg5 it is 7.53 s, and in \citet{2010ApJ...723..267M} 6.2 s. The jet head evolution inside the progenitor is identical, starting sub-relativistic and gradually growing relativistic. For a comparison of this figure see \citet{2010ApJ...723..267M} Fig. 4, and \citet{2000ApJ...531L.119A} Fig 3.

\subsubsection{The three radiative phases}
\label{subsubsec:The three radiative phases}

First we review the properties of radiative phases. As explained in several previous works, collapsar jets show distinct radiative phases \citep[ML07, etc.]{2000ApJ...531L.119A, 2003ApJ...586..356Z, 2007RSPTA.365.1141L, 2011ApJ...740..100B}. The first radiative phase is the ``cocoon breakout'', or the ``precursor phase'', as referred to in ML07. During this phase, the hot cocoon breaks through the stellar surface, forming a quasi-spherical fireball. It is the shortest and least relativistic phase. As outflow is quasi-isotropically spread during this phase, it is the only phase that might be observed at large angles (ML07 \& \citealt{2007RSPTA.365.1141L}). The second radiative phase is the ``shocked phase''. In this phase, the high-pressure cocoon collimates the jet inside the star through multiple tangential shocks, called recollimated shocks. As a consequence, jet has the highest level of collimation and variability during the ``shocked phase'' \citep{2007RSPTA.365.1141L}. Recollimation shocks reflect the strong jet-star interaction during this phase. Around the end of the shocked phase, the last recollimation shock appears. The last recollimation shock marks the limit between inner free-streaming unshocked outflow directly coming from the central engine, and outer shocked outflow. This shock is pushed outward, and after several tens of seconds of engine activity it breaks out, marking the end of the shocked phase. At the same moment, the unshocked outflow starts breaking out almost unperturbed by recollimation shocks, marking the start of the ``unshocked phase'', as a radiative phase. Outflow in this phase is free-streaming; it accelerates according to the adiabatic expansion. The jet gradually gets wider in this phase, after being highly collimated in the previous shocked phase \citep[ML07, etc.]{2007RSPTA.365.1141L}. 

Identifying the start and end of the precursor is simple, but how about the limit between shocked and unshocked phases?  Finding the start of the unshocked phase stands for identifying the breakout of the last recollimation shock \citep[ML07, etc.]{2007RSPTA.365.1141L, 2011ApJ...740..100B, 2013ApJ...777..162M}. At the radius of the last collimation shock, Lorentz factor sharply decreases with radius and pressure increases. This helps identify the limit between the two phases \citep[and ML07]{2009ApJ...699.1261M}. In terms of energy flux, the transition between the first precursor phase and the shocked phase is identifiable as the moment at which the energy flow becomes roughly continuous, although variable. Transition from the shocked to the unshocked phase is the moment at which the on-axis energy drops and becomes steady reflecting the engine constant injection (see Fig 4 in ML07). In our calculations, different from most pervious studies, a relatively deeper injection nozzle is considered, at $10^8$ cm. This deep nozzle adds a dense region to the computation domain, where the star-jet interaction is significantly strong. This affects the development of the unshocked phase's free-streaming core, as the jet is more affected by recollimation shocks, delaying the deployment and breakout of the steady unshocked phase. 

In table \ref{table:breakouts} we summarize our results on breakout times for the different phases (with the corresponding breakout Lorentz factors), for the different calculated models. As in previous studies (ML07 in particular), we confirm the presence of three hydrodynamic phases (for sufficiently long engines). However, as we consider a lower energy injection and a jet birthplace at a relatively deeper region (relative to, e.g., ML07), we have a higher jet-star interaction, and as a result, some additional complexity -- in particular some variability in the unshocked phase. The high jet-star interaction explains the relatively long shocked phases we have, as well as the diversity of this phase from variable to smooth. We believe that the relatively high jet-star interaction in our calculations is more realistic.

\begin{figure*}%[ht] 
    \vspace{4ex}
  \begin{subfigure}%[b]{0.2\linewidth}
    \centering
    \includegraphics[width=0.4\linewidth]{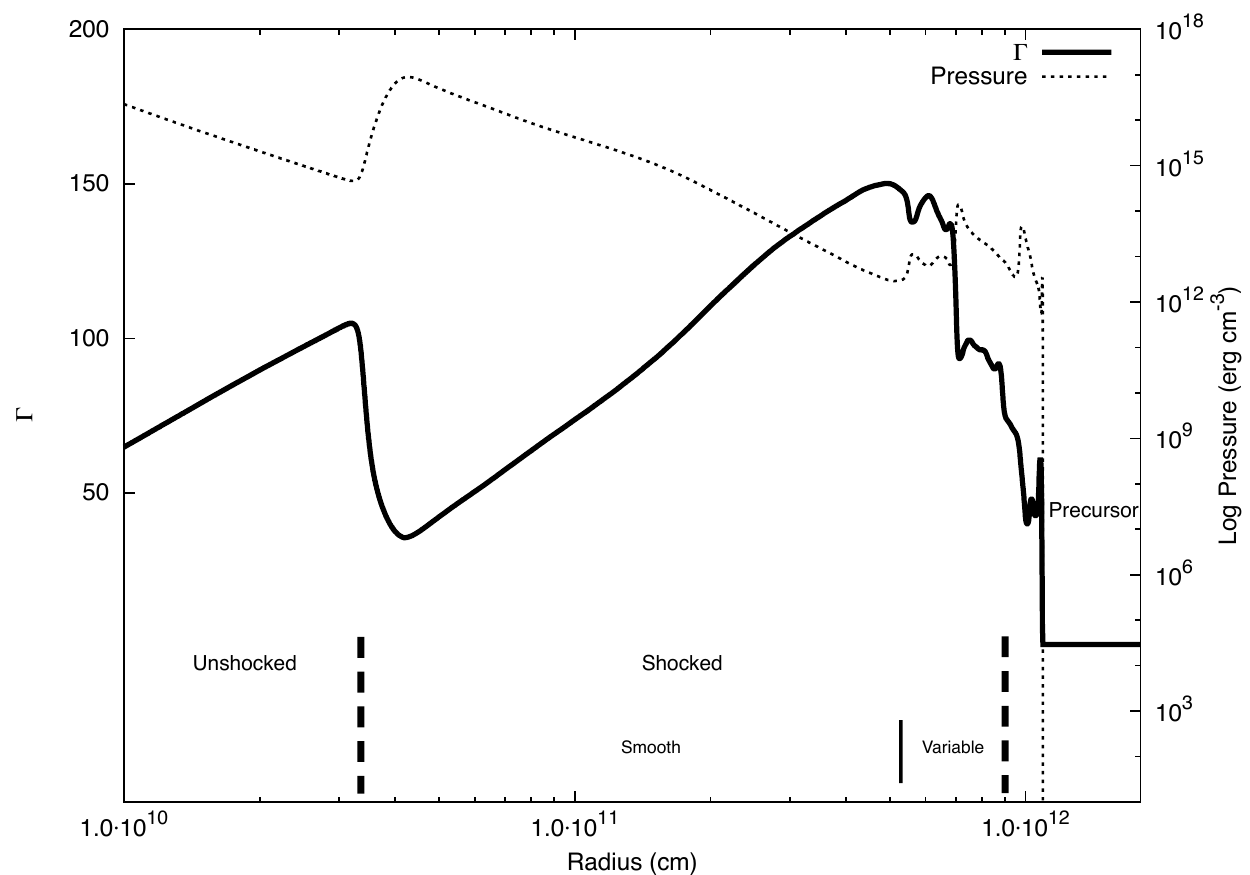} 
    %\vspace{4ex}
  \end{subfigure}%% 
  \begin{subfigure}%[b]{0.2\linewidth}
    \centering
    \includegraphics[width=0.4\linewidth]{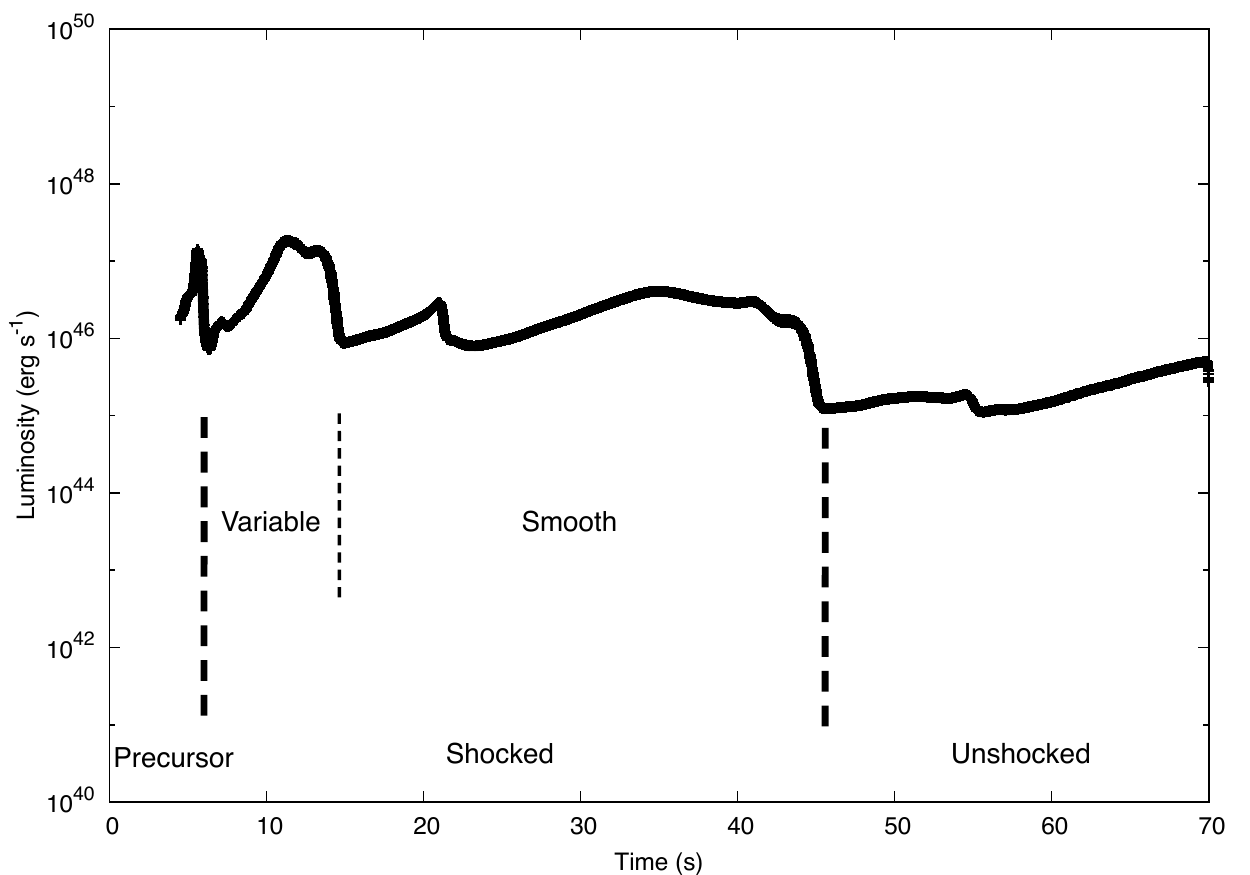} 
    \vspace{4ex}
  \end{subfigure} 
  \caption{On the left, Lorentz factor (solid line) and the pressure (dotted line) along the jet axis as a function of the radius, for L700 model ($T_{inj} = 70$ s), at the moment of the unshocked phase breakout, $t = 40$ s. Dashed lines mark dramatic changes in pressure and $\Gamma$, due to the presence of strong shocks. On the right, the energy flux of the jet measured at $1.2\times10^{11}$ cm, again with the dashed lines separating the three phases.}
  \label{fig:phases} 
\end{figure*}

Figure~\ref{fig:phases} shows the main phases as a function of radius (in the left) and time (in the right) for one long engine in our sample. In the left panel, the first sharp peak in the outer region is the ``precursor'' phase. Next, comes the``shocked'' phase. Shocked phase is very variable, but as it progress throughout the star, it gets progressively smoother, reflecting the steady energy injection. Then, comes the third and last, ``unshocked'' phase. A strong shock separating shocked phase from unshocked phase can be identified on the left panel of figure~\ref{fig:phases} as a sharp decrease in Lorentz factor, and a sharp increase in the pressure, as a function of radius \footnote{Note that the start of the unshocked phase depends of the radius of reference at which measurements are made; after the breakout, fast-unshocked material in the unshocked phase head will continue to catch up to the material at the end of the shocked phase, which affects the length of both phases.}. For a comparison of this figure with previous studies see: \citet{2000ApJ...531L.119A} Fig. 2, \citet{2003ApJ...586..356Z} Figs. 4, 5, and 6, \citet{2006ApJ...651..960M} Figs. 6 and 7, and Fig. 10 in ML07. The right panel of figure~\ref{fig:phases} shows the energy flow along the jet on-axis region, calculated at $1.2\times10^{11}$ cm. The three phases can be identified, and the structure is similar to that in ML07. The unshocked phase, however, is not as steady as in ML07. A comparison of the unshocked phase here, with that of 16TIg5 in ML07 with an injection nozzle at $10^9$ cm and $\sim$5 times more engine energy, shows that we have more variability in our simulations (in particular in the unshocked phase). This extra variability is due to difference in parameters: our inner injection nozzle and lower energy (see Appendix section C for a full comparison). For a comparison of this figure see: ML07 Fig. 4.

%%%

%%%

\subsubsection{Breakout times}
\label{subsubsec:Breakout times}

\begin{figure*}%[ht] 
    \vspace{4ex}
  \begin{subfigure}%[b]{0.2\linewidth}
    \centering
    \includegraphics[width=0.4\linewidth]{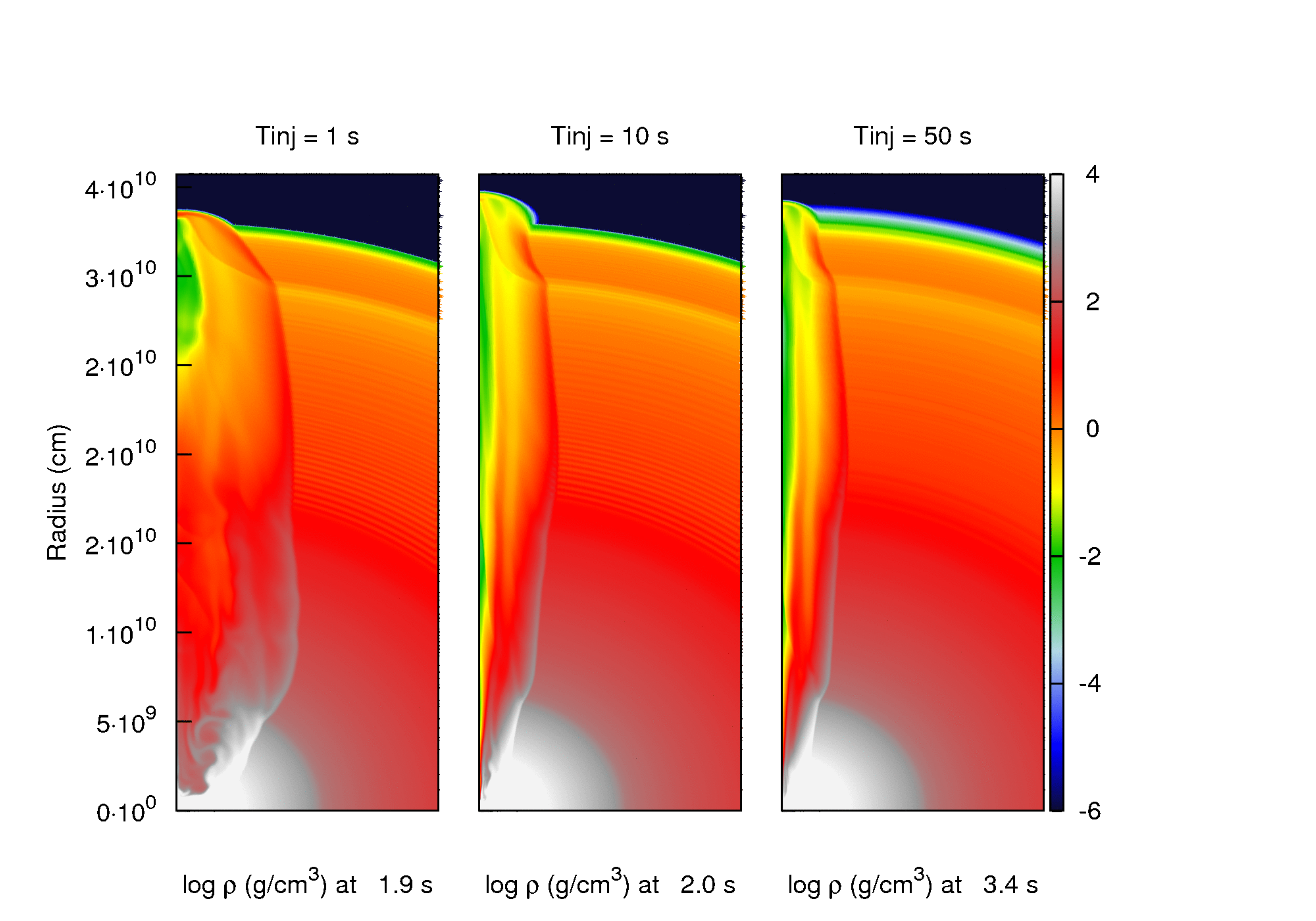} 
    %\vspace{4ex}
  \end{subfigure}%% 
  \begin{subfigure}%[b]{0.2\linewidth}
    \centering
    \includegraphics[width=0.4\linewidth]{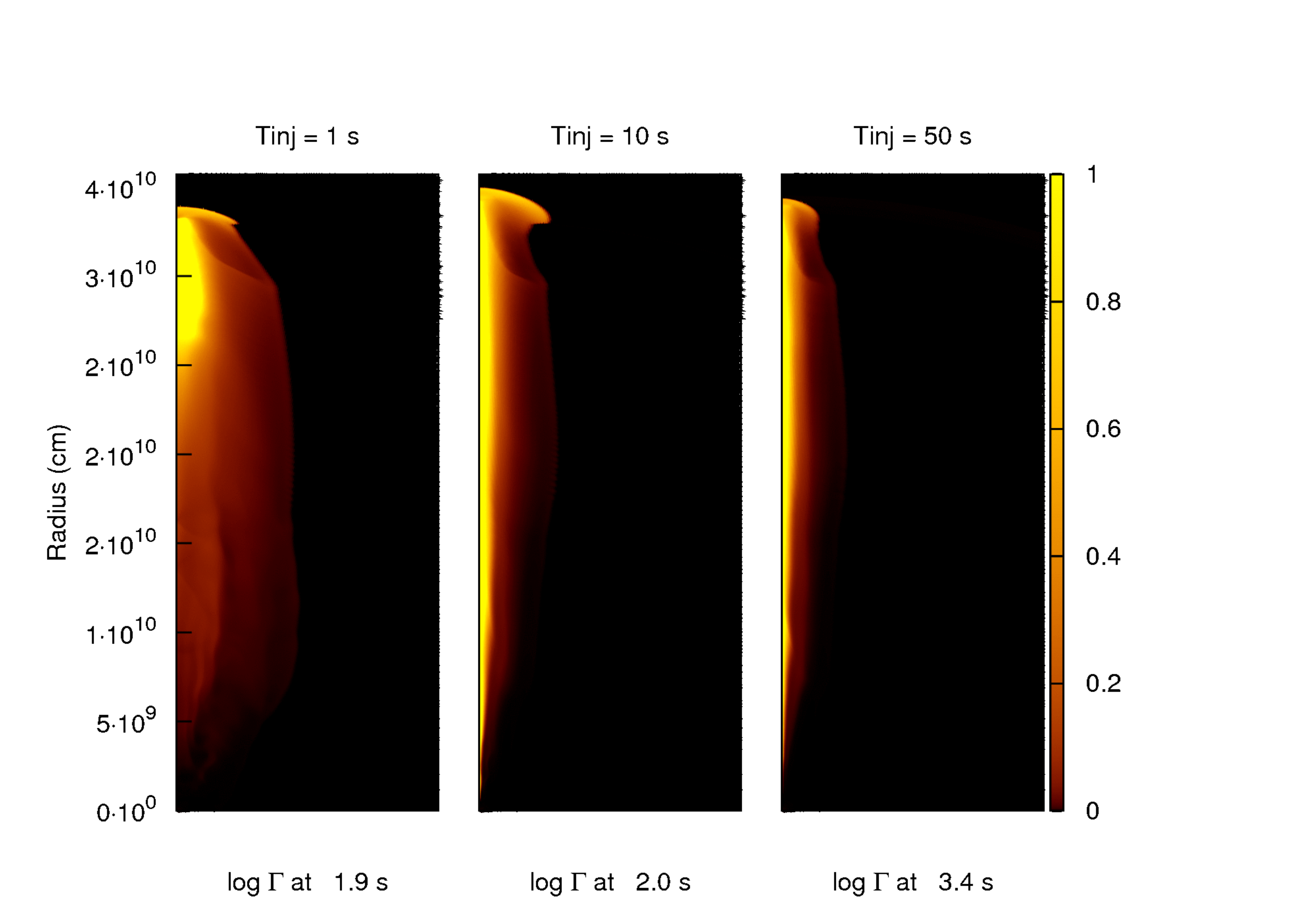} 
    %\vspace{4ex}
  \end{subfigure} 
  \begin{subfigure}%[b]{0.2\linewidth}
    \centering
    \includegraphics[width=0.4\linewidth]{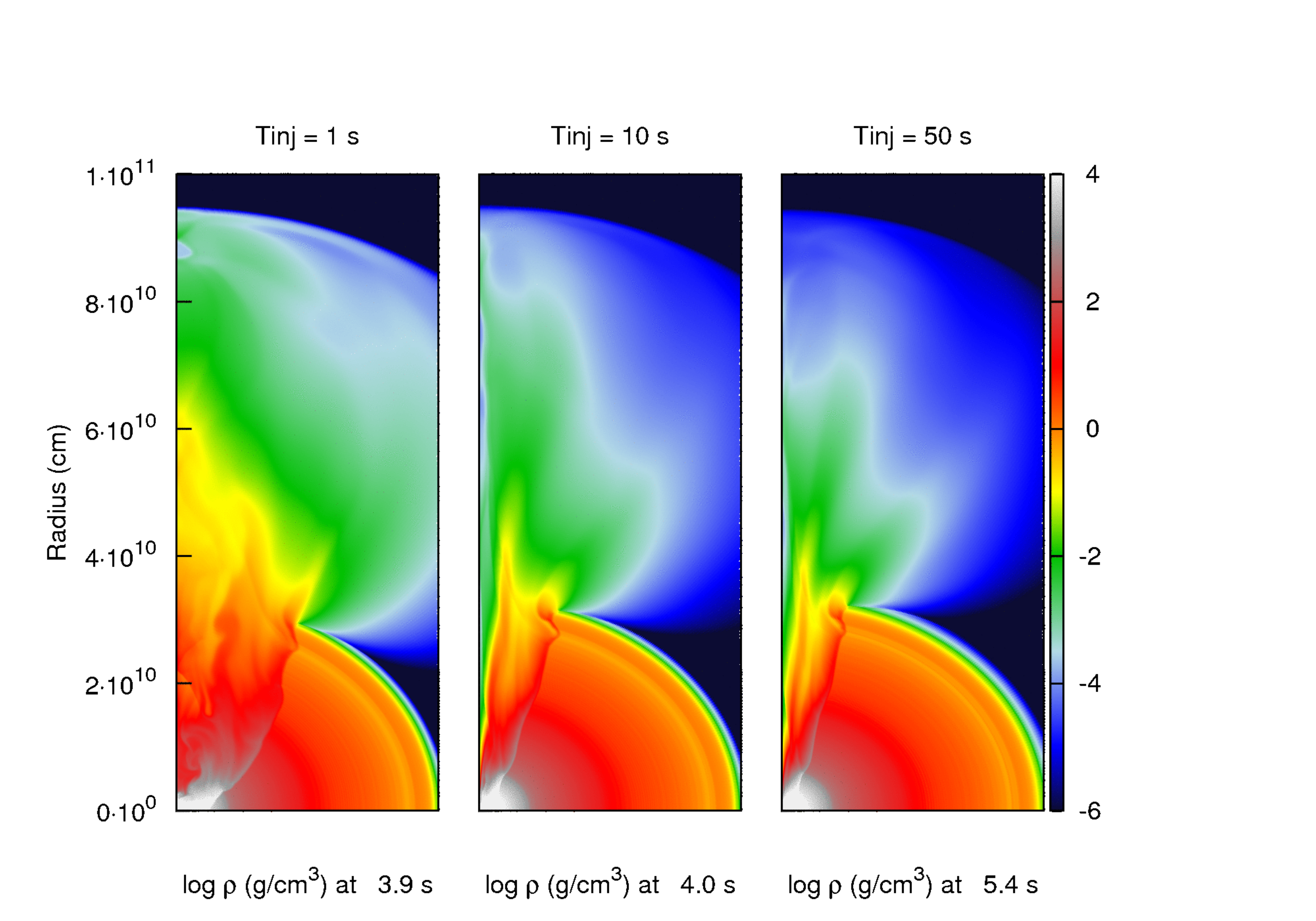} 
  \end{subfigure}%%
  \begin{subfigure}%[b]{0.2\linewidth}
    \centering
    \includegraphics[width=0.4\linewidth]{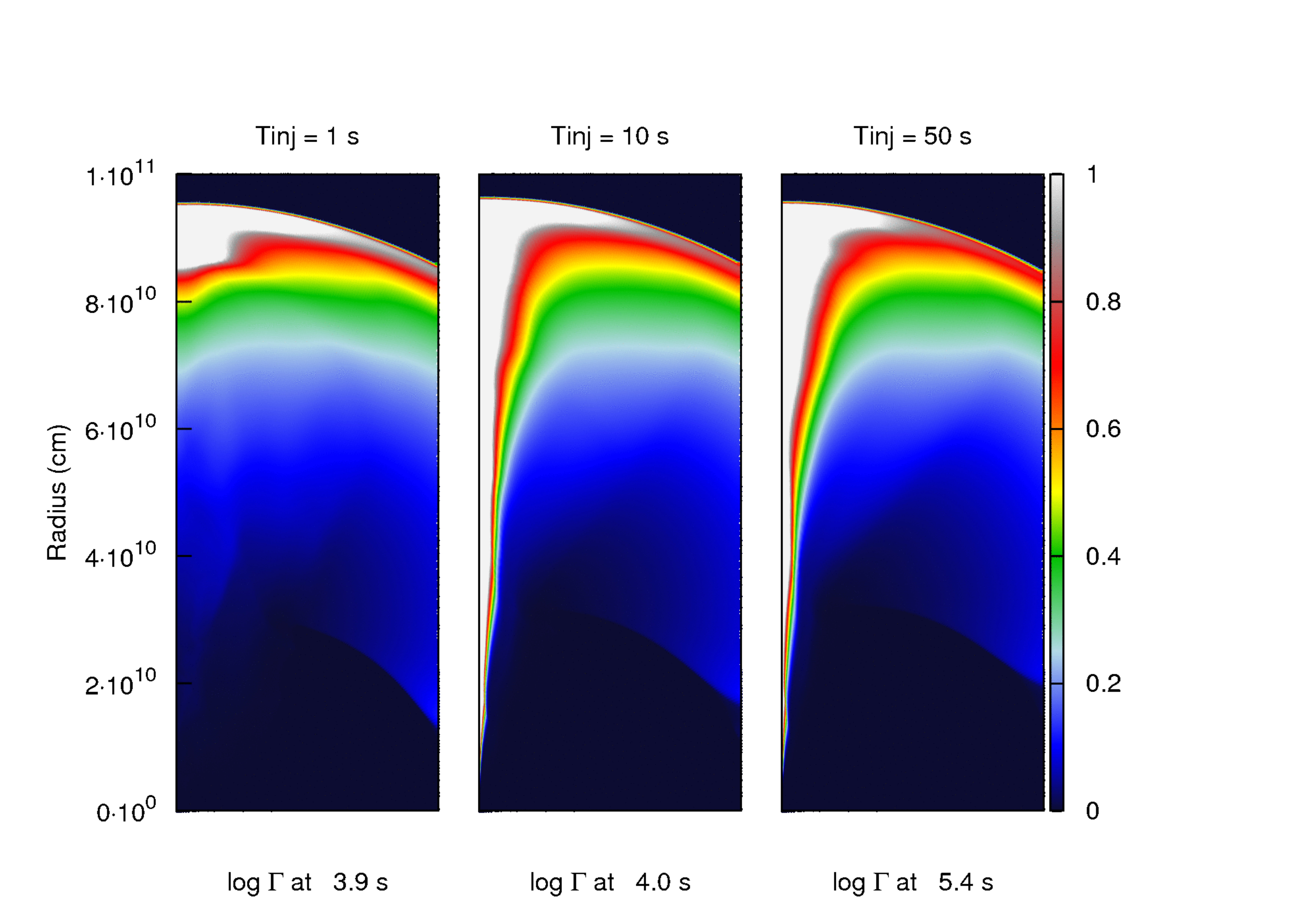} 
  \end{subfigure} 
  \caption{Density (at the left) and Lorentz factor (at the right), at the cocoon breakout time (in the top) and 2 s after (bottom). Breakout density and Lorentz factor (top panels) reveal a wide jet for the brief engine model (1 s) and a well-collimated jet for the two other longer engines (10 \& 50 s). After the breakout (bottom panels), dense materials and a poorly collimated jet is visible for the brief engine, while longer engines efficiently produce a clear jet structure.}
  \label{fig:snapshots} 
\end{figure*}

In figure~\ref{fig:snapshots} we show snapshots of the density and Lorentz factor at, and 2 s after the cocoon breakout, in the top two and bottom two panels, respectively, for a brief, intermediate and a long engine (from left to right, respectively). There is a contrast between brief engine's jet and the other engines' jets. First, in the brief engine model, the outflow shows significant sideway expansion, during and after the cocoon breakout. While in the other models, a much more collimated and typical jet structure is visible. Second, density is significantly higher in brief engine model, in particular that of the jet head (1-2 orders of magnitude higher). The tendency of the jet head getting denser can be found in equation (\ref{eq:den}). This contrast explains why brief engines have not been suggested as favorable to power GRBs in the relativistic/collimated jet scenario (and thus less studied), whereas the previously cited numerous and deep studies attempting to reproduce extreme GRBs with long engine models.

\begin{figure}
 \includegraphics[width=\columnwidth]{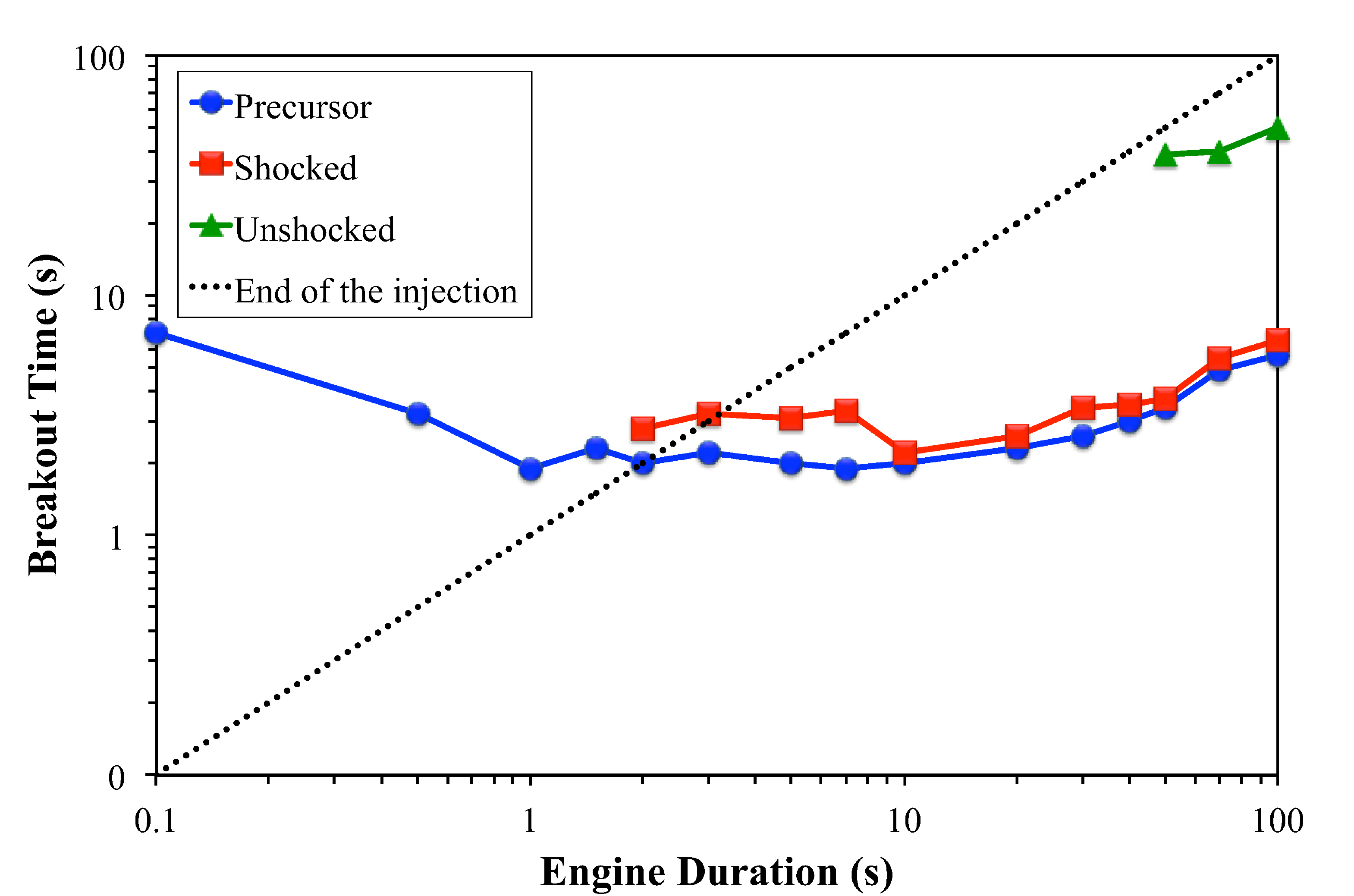}
 \caption{Breakout times, for the radiative phases of the computed models: precursor in blue, shocked in red, and unshocked in green. The dotted line indicates the time when the injection stops for each model.}
 \label{fig:breakout times}
\end{figure}

Next we focus on the breakout time as a function of engine duration. Previously \citet{2012ApJ...750...68L} studied the cocoon breakout times for engine durations from 2 to 15 s (see Figure 4 in \citealt{2012ApJ...750...68L}). Here, we extend the same study to a much wider engine duration range (0.1 to 100 s). Furthermore, we cover breakout times of all the three radiative phases. Results are shown in figure~\ref{fig:breakout times}. 

For the first group, brief engines, launched jets show only the cocoon phase. Also, in brief engines, cocoon breakout times are relatively long, longer than the engine duration. For all other longer engines, the engine is still running during, and after, the breakout (see the dotted line in figure~\ref{fig:breakout times}). This element will have profound consequences on brief engines' jets performance, relative to the other longer engines' jets. 

In Short and intermediate engine models, jet head expansions are the fastest in our sample, and breakout times are the shortest. Both short and intermediate engines produce a jet consisting of a precursor followed by the shocked phase. After the shocked phase, comes the steady unshocked phase. As our injection nozzle is relatively deep, the star-jet interaction is stronger, than in ML07, and thus the relatively longer shocked phase. Only long engines are long enough to have their jets include the unshocked phase at the end. Breakout times for the three phases are relatively later for this group of long engines. This is due to the long engines low energy deposition rate, inversely proportional to the engine duration (as we assume the same injected total energy for all engines). Thus, long engines' jets are softer ($10^{49-50}$ erg s$^{-1}$), and need more time to accumulate energy, enough to buildup a strong relativistic shock.

The analyze of the radiative phases leads to findings similar to those in \citet{2012ApJ...750...68L}. In summary: i) breakout times are the shortest in short and intermediate engine models, while having a tendency of getting longer in the two limits (brief and long engine duration models); and ii) the total combination of the three radiative phases suggests that short and intermediate engines produce the most variable jets, thanks to the high contribution of the shocked phase and the absence of the smooth unshocked phase in their jets. Note that high variability is one key feature observed in the light curve of many GRBs, and that it implies either a variable engine or a variable jet.

\begin{table}
 \caption{Breakout times of the main phases at the star surface (in the on-axis region).}
 \label{table:breakouts}
 \begin{tabular}{lcccc}
  \hline
& Precursor & \multicolumn{2}{c}{Shocked} & Unshocked\\[0.5pt]
& Breakout (s) & \multicolumn{2}{c}{Breakout (s)} & Breakout (s)\\[0.5pt]
Model & ($\Gamma$) & \multicolumn{2}{c}{($\Gamma$)} & ($\Gamma$)\\[0.5pt]
  \hline
  B001 & 7.0 (1.02) & - & - & -\\
B005 & 3.2 (1.1) & - & - & -\\
B010 & 1.9 (33) &- &- &-\\
B015	 & 2.3 (37) & - & - & -\\
S020 & 2.0 (34) & 2.8 (52) & - & -\\
S030 & 2.2 (43) & 3.2 (70) & - & -\\
S050	 & 2.0 (40) & 3.1 (55) & - & -\\
S070	 & 1.9 (34) & 3.3 (65) & 5.7 (68)$^1$ & -\\
I100	& 2.0 (37) & 2.2 (42) & 8.2 (77)	& -\\
I200 & 2.3 (25) & 2.6 (23) & 8.0 (75) & -\\
I300 & 2.6 (29) & 3.4 (36) & 7.3 (62) & -\\
I400 & 3.0 (17) & 3.5 (30) & 6.4 (59) & -\\
L500 & 3.4 (15) & 3.7 (26) & 7.0 (55) & 38.9 (77)$^1$\\
L700	 & 4.9 (14) & 5.5 (22) & 9.0 (44) & 40.1 (104)\\
L999	 & 5.6 (14) & 6.5 (32) & 10.2 (31) & 50.1 (109)\\
16TIg5$^2$ & 6.8 (20) & 7.8 (40) & - & 25.0 (71)\\
  \hline
 \end{tabular}
 \vspace{1ex}
 
     \raggedright {\footnotesize $^1$ Not powered long enough to take part in the breaking out jet or influence it significantly.}
     
     \raggedright {\footnotesize $^2$ A model computed using the same injection nozzle as in ML07, $10^9$ cm, instead of the $10^8$ cm used for all the above models. The same total energy used in ML07 ($5.32\times10^{52}$ erg), and thus the same engine luminosity, was also considered.}
\end{table}

\subsection{Light curve and variability}
\label{subsec:Light curve and variability}

Table \ref{table:LCs} shows possibility of relativistic outflow contribution in the jet for the different radiative phases, and for each of our models. The different contributions would provide information on the jet light curve and its features, based on the nature of the contributing phases and the rate of the contribution. In our analyze we are limited to on-axis observation, however as GRBs are observed with line of sights very near to the on-axis region as well (as their huge brightness and energy output suggest), a limitation to the on-axis region is fine as long as we compare to typical GRBs. 

From table \ref{table:LCs}, one naive judgment based on jet variability alone (high in a typical GRB light curve) on whether a GRB could be observed, is that jets rich in variability are necessary. In our sample, short, intermediate, and to a lesser extent long engines seem to provide variability. 

\begin{figure*}%[ht] 
    \vspace{4ex}
  \begin{subfigure}%[b]{0.2\linewidth}
    \centering
    \includegraphics[width=0.4\linewidth]{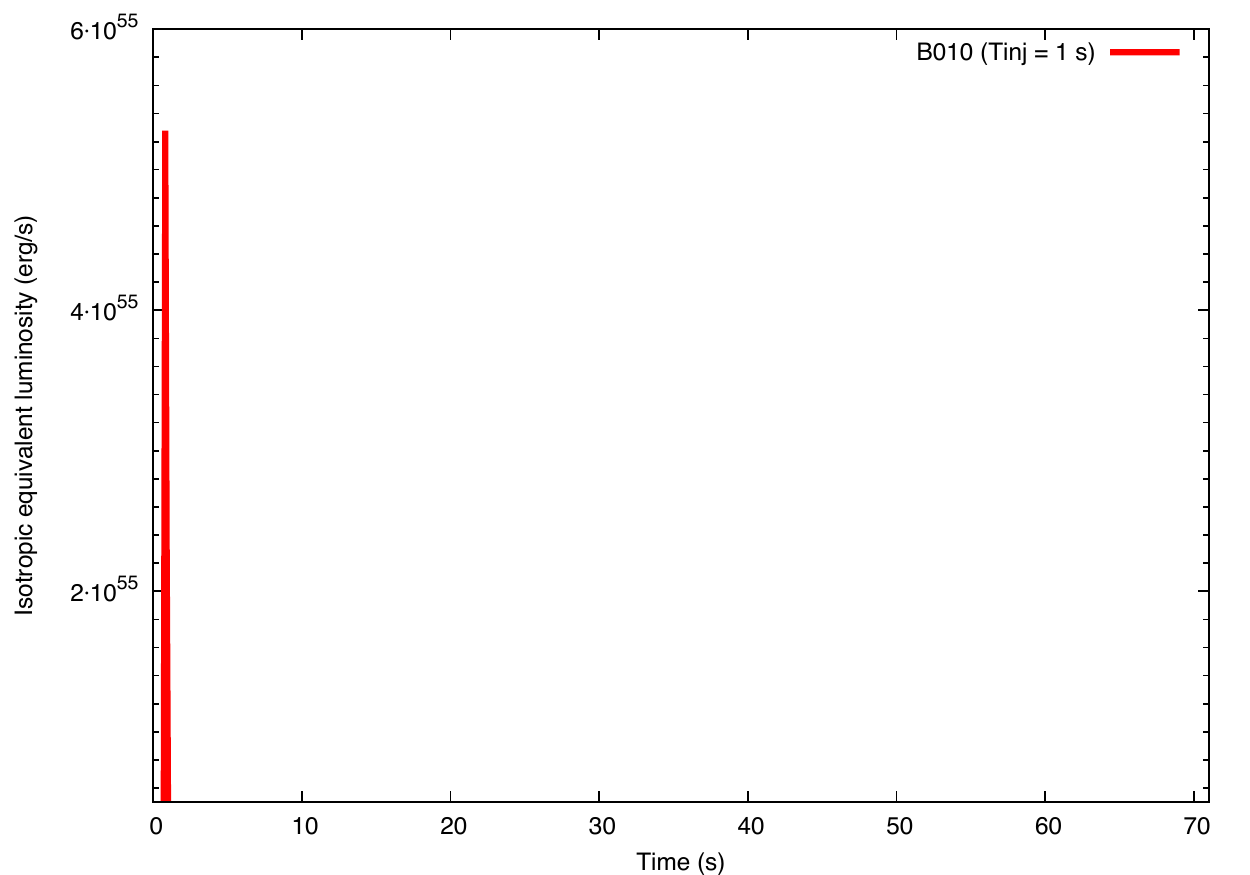} 
    %\vspace{4ex}
  \end{subfigure}%% 
  \begin{subfigure}%[b]{0.2\linewidth}
    \centering
    \includegraphics[width=0.4\linewidth]{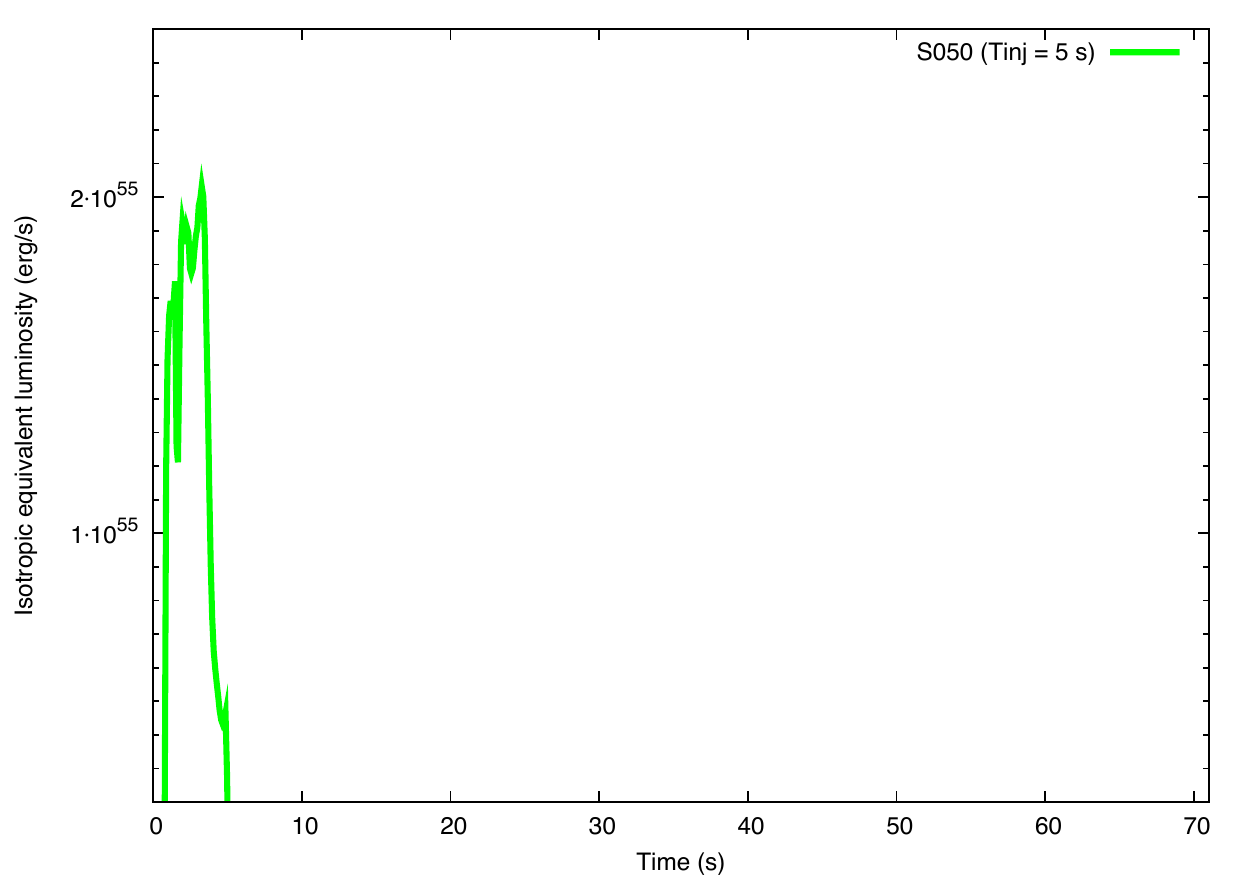} 
    %\vspace{4ex}
  \end{subfigure} 
  \begin{subfigure}%[b]{0.2\linewidth}
    \centering
    \includegraphics[width=0.4\linewidth]{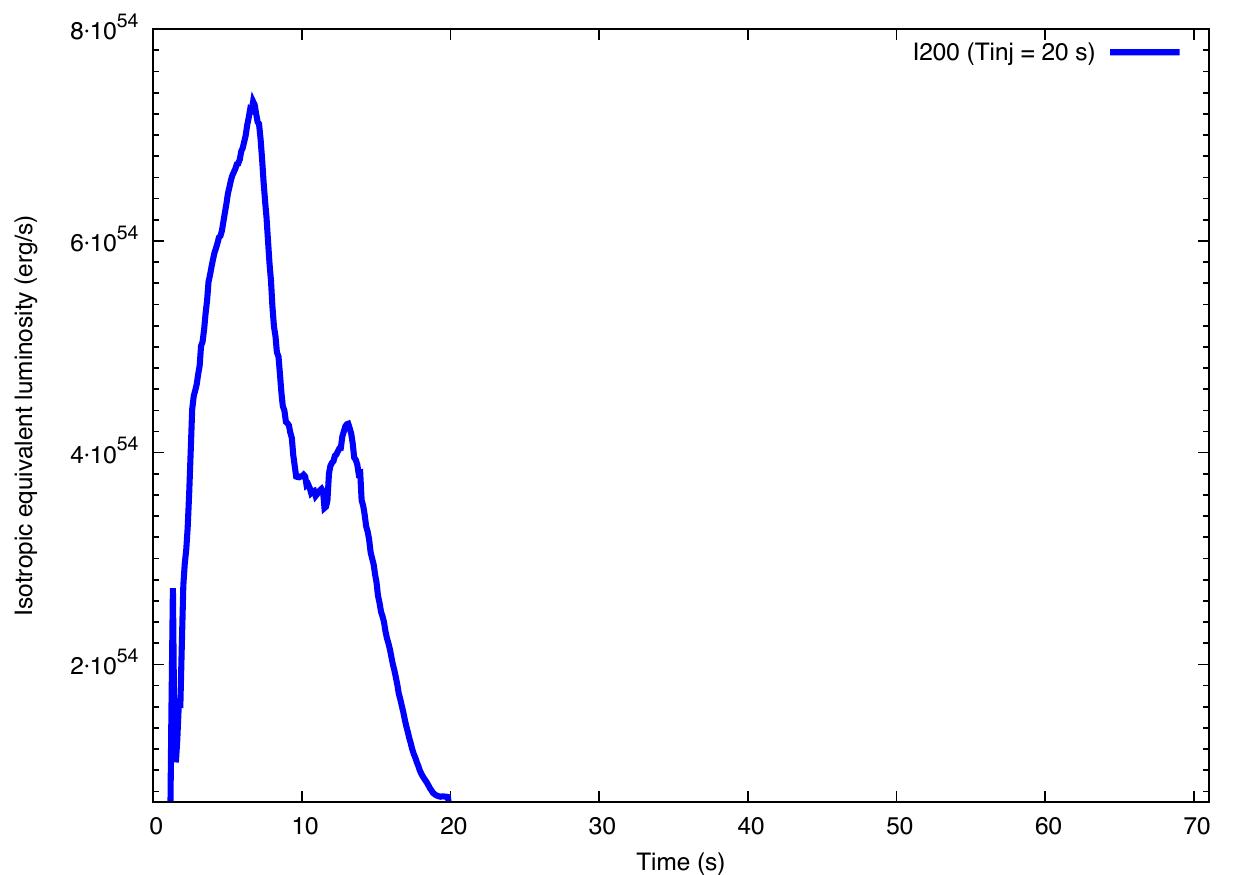} 
  \end{subfigure}%%
  \begin{subfigure}%[b]{0.2\linewidth}
    \centering
    \includegraphics[width=0.4\linewidth]{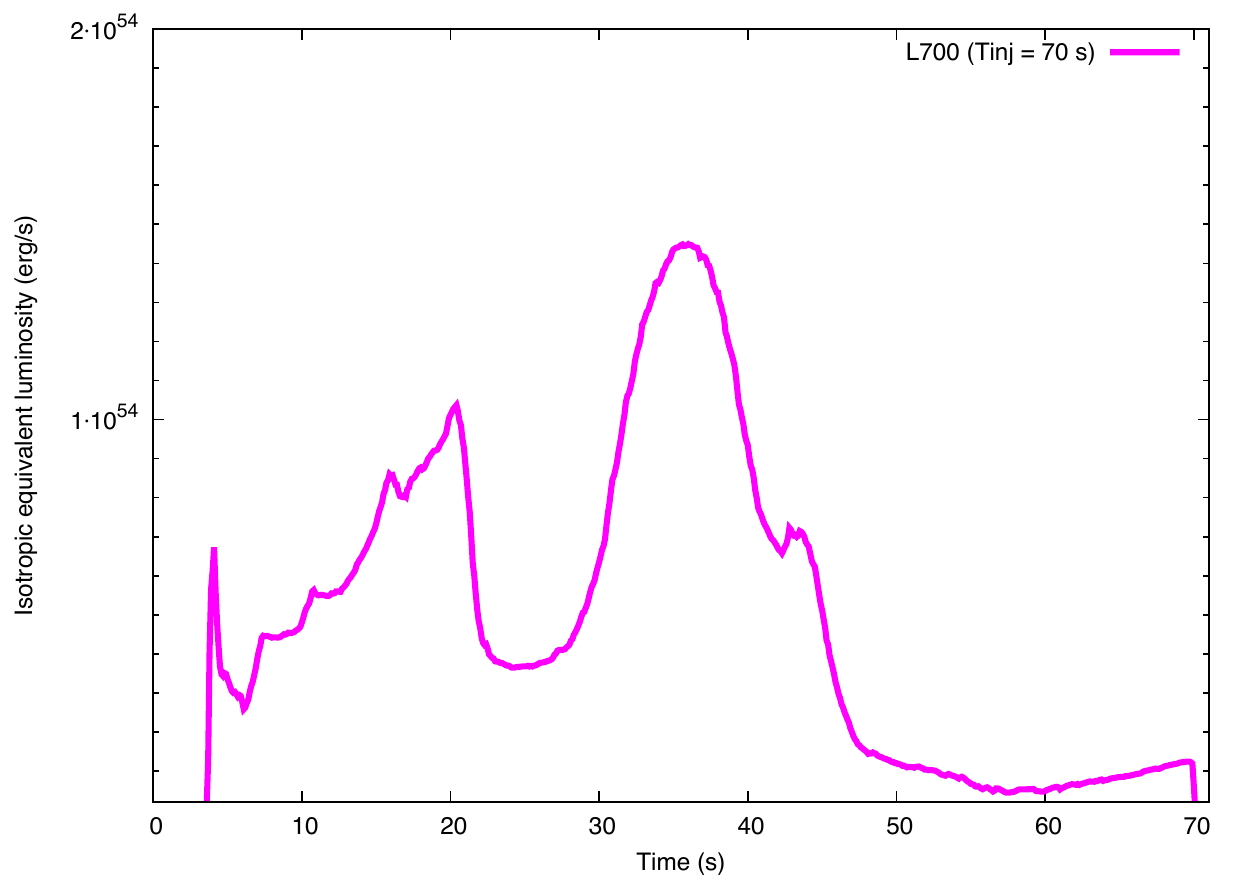} 
  \end{subfigure} 
  \caption{Light curves from on-axis observer of the four different types of engine models. On the top left, for brief engine jet ($T_{inj} = 1$ s), the light curve is a single sharp peak. On the top right, for a short engine jet ($T_{inj} = 5$ s) the peak is wider and shows high variability. On the bottom left, for an intermediate engine ($T_{inj} = 20$ s) the light curve display two bulks structure. Finally, for a long engine the jet's light curve ($T_{inj} = 70$ s) show more that two bulks, a more complex structure. Jet light curves were estimated as explained in \S \ref{sec:Data processing procedure}, and after ML07.}
  \label{fig:LCs} 
\end{figure*}

To have a closer look, we estimate light curves following the very same method used by ML07 (Fig. 12 \& \S 4.1 of ML07 for details). We measure energy at the same location ($1.2\times10^{11}$ cm), with the same condition on the outflow ($\Gamma$ > 10), with the same assumptions, and with the same angular bins (as explained in \S \ref{sec:Data processing procedure}). ML07 argued that such light curves would be a good proxy to evaluate prompt emission light curves. Figure~\ref{fig:LCs} shows four light curves representing our four classes of engines. We assume an observer line of sight in the on-axis region. We calculate isotropic equivalent luminosities (as in ML07). 

Brief engines produce a sharply single peaked light curve; it is the signature of the mildly relativistic precursor phase, a breakout shock. Short engines produce a longer and very variable single peaked structure. This is due to the large contribution of the shocked phase dominating short engines' jets. Intermediate engines are long enough to display a second bulk (I200 in figure~\ref{fig:LCs}, from $\sim$10 to 20 s), showing a double peaked structure. Intermediate engines' jets as well deploy almost all the engine energy in the form of a shocked phase, producing high variability. Long engines are long enough to have a significant contribution from the smooth unshocked phase. In this last phase, the luminosity sharply decreases showing a steady and smooth evolution that reflects the steadily injecting engine (L700 in figure~\ref{fig:LCs}, from $\sim$47 to 70 s). With this last phase, the structure of such long engines is the more complex, resulting from the contribution of all the three different phases, however it's less variable than in the case of intermediate engines, especially at the end.

Note that isotropic equivalent luminosities in light curves of figure~\ref{fig:LCs} are very high, since we considered an observer line of sight very close to the on-axis, at $0.125^{\circ}$ (which gives an isotropic factor of $\sim10^5$). Also, these light curves take into account all the jet relativistic energy (a radiative efficiency of 100\% for relativistic outflow $\Gamma$ > 10). The observed luminosities and timescale will differ. However, the observed prompt emission's light curve is expected to trace these light curves (in particular the degree variability), as the considered outflow is relativistic (ML07).

\begin{table*}
 \caption{Light curve structure according to the contribution of the jet main phases (considering the energy flux at $1.2\times10^{11}$ cm for an on-axis observer and outflow with $\Gamma$ > 10, as in ML07).}
 \label{table:LCs}
 \begin{tabular}{ccccccc}
  \hline
& Precursor Phase & \multicolumn{2}{c}{Shocked Phase} & Unshocked Phase & &\\
& (Quasi-Isotropic) & \multicolumn{2}{c}{(Well-Collimated)} & (Collimated) & & Jet light curve\\
    Model & Breakout & Variable & Smooth & Steady & Duration$^{*}$ & Structure\\
  \hline
    B001 & Yes & - & - & - & - & Sharp Narrow Peak\\
    B005 & Yes & - & - & - & - & Sharp Narrow Peak\\
    B010 & Yes & - & - & - & 0.4 & Sharp Narrow Peak\\
    B015 & Yes & - & - & - & 0.6 & Sharp Narrow Peak\\
    S020 & Yes & Yes & - & - & 1.4 & Wide Variable Peak\\
    S030 & Yes & Yes & - & - & 2.2 & Wide Variable Peak\\
    S050 & Yes & Yes & - & - & 4.4 & Wide Variable Peak\\
    S070 & Yes & Yes & - & - & 6.5 & Wide Variable Peak\\
    I100 & Yes & Yes & $\Delta$ & - & 9.4 & Wide Variable Peak\\
    I200 & Yes & Yes & Yes & - & 19.1 & Two Main Bulks\\
    I300 & Yes & Yes & Yes & - & 28.8 & Two Main Bulks\\
    I400 & Yes & Yes & Yes & - & 38.5 & Two Main Bulks\\
    L500 & Yes & Yes & Yes & $\Delta$ & 48.0 & More than Two Bulks\\
    L700 & Yes & Yes & Yes & Yes & 66.6 & More than Two Bulks\\
    L999 & Yes & Yes & Yes & Yes & 95.7 & More than Two Bulks\\
  \hline
  \multicolumn{7}{l}{$\Delta$: This last phase, or part, is launched, but not long enough to contribute in the light curve.}\\
  \multicolumn{7}{}{l}{$^*$: Time during which outflow has been measured. This timescale varies strongly with the angle. Note that it is not easily linked to $T_{90}$,}\\ 
  \multicolumn{7}{}{l}{especially for chocked jets; as one requirement for the two timescales to converge is that outflow must be highly relativistic.}\\

 \end{tabular}
\end{table*}

%%%%%%%%%%%

\subsection{Angular structure and collimation}
\label{subsec:Angular structure and collimation}

A collimated outflow, in a jet structure, is an essential requirement to explain typical GRBs (e.g., extreme isotropic equivalent energy; \citealt{2000PhR...333..529P}). In this section we study the degree of collimation for jets launched by the different engine duration models. Then we estimate the favorability of the different engines durations, at producing a typical GRB.

\begin{figure}
 \includegraphics[width=\columnwidth]{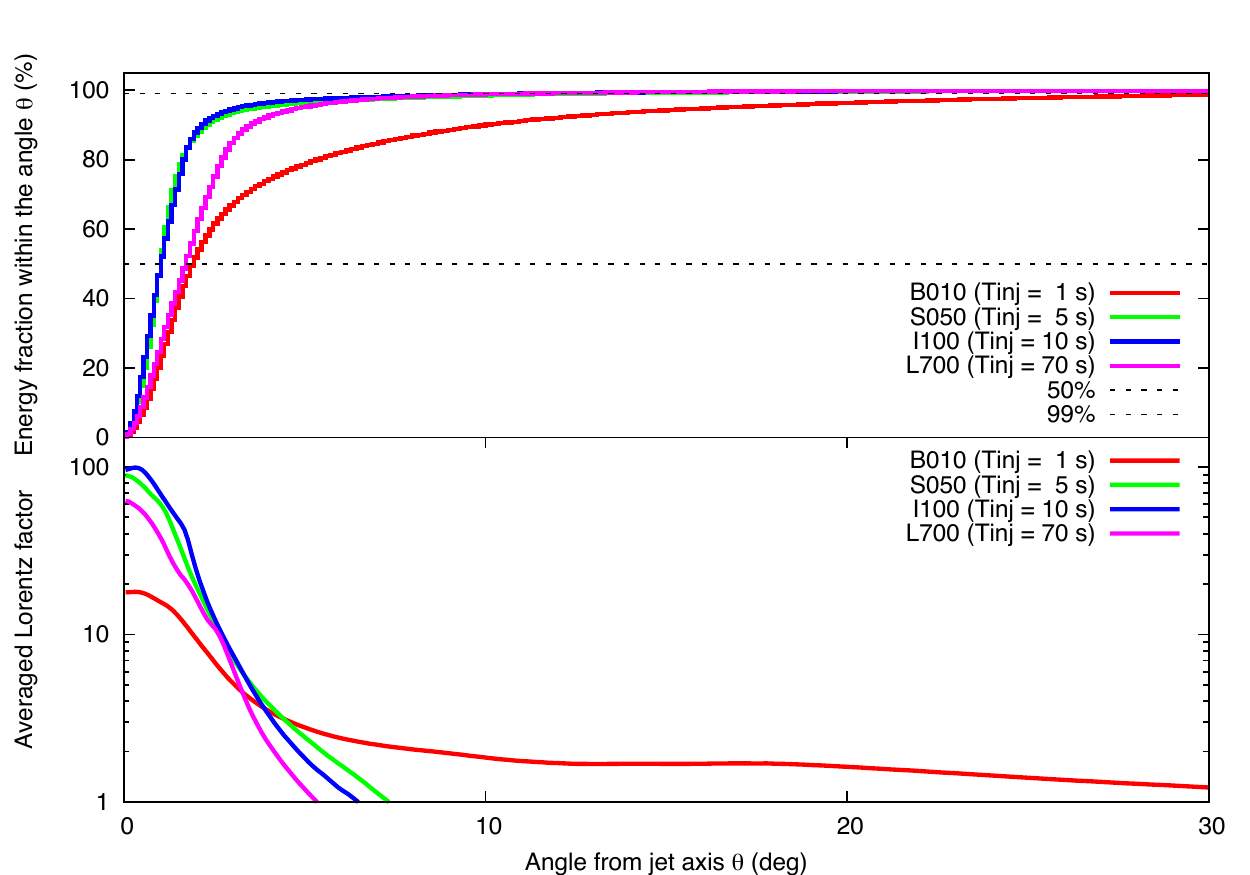}
 \caption{On the top panel, the cumulative distribution of the jet energy fraction inside the angle $\theta$, with angle (brief engine in red, short in green, intermediate in blue and long in cyan). The two dashed lines help identify angles inside which 50\% and 99\% of the energy is contained, and how these angles differ for the four engine models. The bottom panel shows ``the averaged Lorentz factor'' as a function of the angle. It is calculated via the ratio of energy to mass at the given angle (same as in \citealt{2015ApJ...813...64D}).}
 \label{fig:Gamma av}
\end{figure}

\begin{figure}
 \includegraphics[width=\columnwidth]{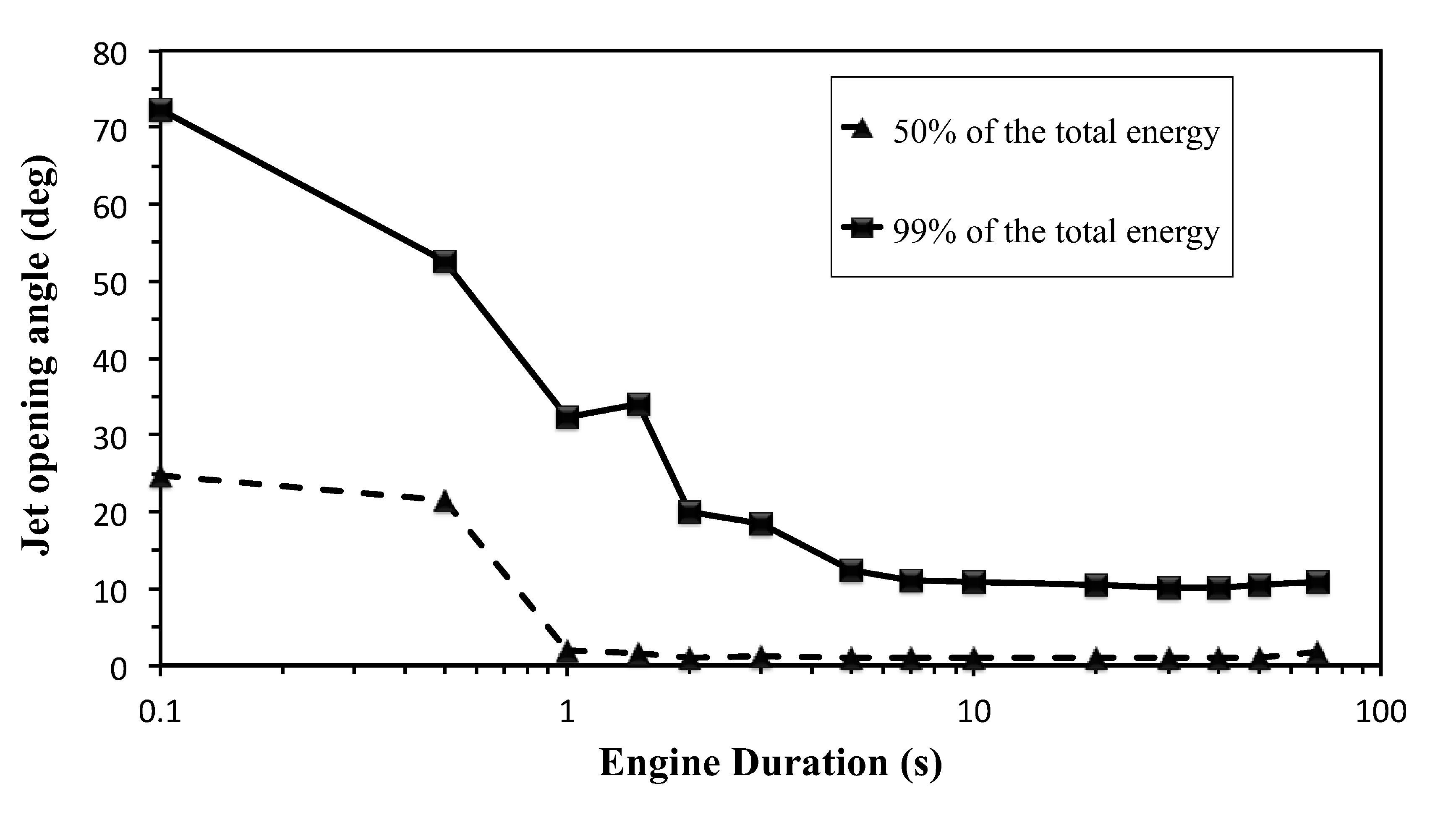}
 \caption{The opening angle of outflow containing 50\% of the total energy $\theta_{50}$, in triangles with a dashed line, and outflow containing 99\% of the total energy $\theta_{99}$, in squares and a solid line (for more information see figure \ref{fig:Gamma av}).}
 \label{fig:OAs}
\end{figure}

In top panel of figure~\ref{fig:Gamma av}, we show the cumulative distribution of the jet total energy, inside the polar angle $\theta$, as a fonction of the same angle $\theta$. Two horizantal lines help identify the angle inside which half, and nearly all the energy (99\%) carried by the jet is contained. In figure~\ref{fig:OAs}, we explicitly show these two angles, and how they vary for the different engine durations (in dashed blue, $\theta_{50}$ and in solid red, $\theta_{99}$, respectively).

Our results illustrate a clear evolution, the longer the engine duration is, the more concentrated (in the on-axis region) the energy is (up to intermediate engines domain where the concentration is maximized). In figure~\ref{fig:Gamma av} (bottom) we present a closely related parameter to angular distribution and collimation, which is Lorentz factor (averaged for all outflow, calculated from the ratio of energy to mass, at the given angle; \citealt*{2015ApJ...813...64D}). For the brief engine model, the outflow is, on average, relativistic (averaged Lorentz factor $\sim$20) at the on-axis region, and mildly relativistic at larger angles ($\sim$2 to 3). This is in total contrast with the other longer engines being highly relativistic ($\sim$100) and much less relativistic (< 2), at large angles in the off-axis region. The intermediate engine model; closely followed the short engine model; produce the highest averaged Lorentz factors; while long engine model produce less relativistic outflow at all angles. 

Combing the energetic and relativistic properties, we can understand more about the angular structure of the different models. Brief engines are contrastive with longer engines, in terms of relativistic and angular structure. Brief engines produce failed jets, mildly relativistic outflow, poorly collimated and spared on large angles. While for the longer engines, outflow is highly relativistic, and concentrated in a few degrees near the on-axis region, that is, a typical ``jet'' structure (although the degree of collimation seems to vary).

This result of failed/successful jet for short/long engine duration is not new, as \citet{2012ApJ...750...68L} found the same from a different parameter, the eject velocity ($\beta\Gamma$). However, our study goes further as it's the first at exploring the energetic, relativistic and angular distribution. It also covers a much larger engine duration domain. What is surprising however, is the behavior at the other extreme, at long engine durations (50-100s).

We note that our study on angular distribution (figure \ref{fig:Gamma av} \& \ref{fig:OAs}) helps to link to one other major study. \citet*{2013ApJ...777..162M}, using a few tens seconds engines, suggested that the naive prediction on the jet opening angle $\theta_{jet}\sim$1/$\Gamma_0$ ($\sim12^{\circ}$ for $\Gamma_0$ = 5) is rough, and proposed that $\theta_{jet}\sim$0.2/$\Gamma_0$ ($\sim2-3^{\circ}$) is a better alternative. They explained that at $\theta_{jet} >$ 0.2/$\Gamma_0$, there is no jet outflow (although there is some energy), but instead a baryon-rich sheath with $\Gamma_s$ < 5. Our widely diverse simulations represent a good laboratory to test their formula. We found that their suggestion is indeed an excellent alternative to estimate the opening angle, over a large duration domain (from short to long engines). As it can be seen at the bottom panel of figure~\ref{fig:Gamma av} the true $\theta_{jet}$ of the jet is $\sim2-3^{\circ}$ ($\sim$0.2/$\Gamma_0$) and for $\theta_{jet} >$ 0.2/$\Gamma_0$ there is a baryon-rich sheath with $\Gamma_s < 5$, exactly as suggested in \citet{2013ApJ...777..162M}! Furthermore, our study adds a deeper reading on $\theta_{jet}$. Our results show that the new prediction cannot be generalized to all types of engine, as brief engines $T_{inj} \ll T_{breakout}$ (e.g. B001 \& B005) are outliers. This is due to the fact that these jets contain one phase (the cocoon phase) that gets baryon loaded from the baryon-rich stellar envelope, reducing its breakout Lorentz factor to $\ll 5\Gamma_0$. To summarize, jet opening angle can be writen in a more general picture: For a successfully launched collapsar jet, $\theta_{jet}\sim$0.2/$\Gamma_0$; while for a failed/chocked jet, $\theta_{jet} >$ 0.2/$\Gamma_0$.

Although figure \ref{fig:Gamma av} and \ref{fig:OAs} show the contrast between failed and successful jets, a deeper comparison of jet collimation and how it varies in long engine durations is necessary. We carry out a quantitative comparison between each engine's on-axis and off-axis energetic and relativistic properties, in order to visualize the degree of collimation. Since our relativistic jets are of the standard type (as defined in \citealt*{2002NewA....7..197R}, see figure 1-a; same as ``power-law universal'' definition in \citealt*{2004NewAR..48..459L};\citealt*{2005ApJ...620..355L}), an on-axis to off-axis comparison (i.e., a ratio) allows to quantitatively visualizing the degree of collimation along the on-axis. We use the same data, angular bins as explained in \S \ref{sec:Data processing procedure}, and at the same radius ($1.2\times10^{11}$ cm). We consider all energy, including outflow at both relativistic and non-relativistic velocities. By on-axis region, we imply the innermost angular bin centered at $0.125^{\circ}$; while for off-axis region we consider the 21st angular bin, centered at $10^{\circ}$ away from the jet axis. Our choice of an off-axis region at $10^{\circ}$ is because it is the same opening angle used for energy injection at the nozzle, and thus $\sim10^{\circ}$ would be the edge of a fully evolved jet. Effectively, it's the farthest angle from the on-axis at which outflow from all the models is present, thus allowing the comparison to be made for all our engine models. We analyze the averaged Lorentz factor, and the total energy in both on-axis and off-axis regions. Finally, we compare on-axis / off-axis ratios, for the energy and Lorentz factor, for the different engine models. 

\begin{figure*}%[ht] 
    \vspace{4ex}
  \begin{subfigure}%[b]{0.2\linewidth}
    \centering
    \includegraphics[width=0.4\linewidth]{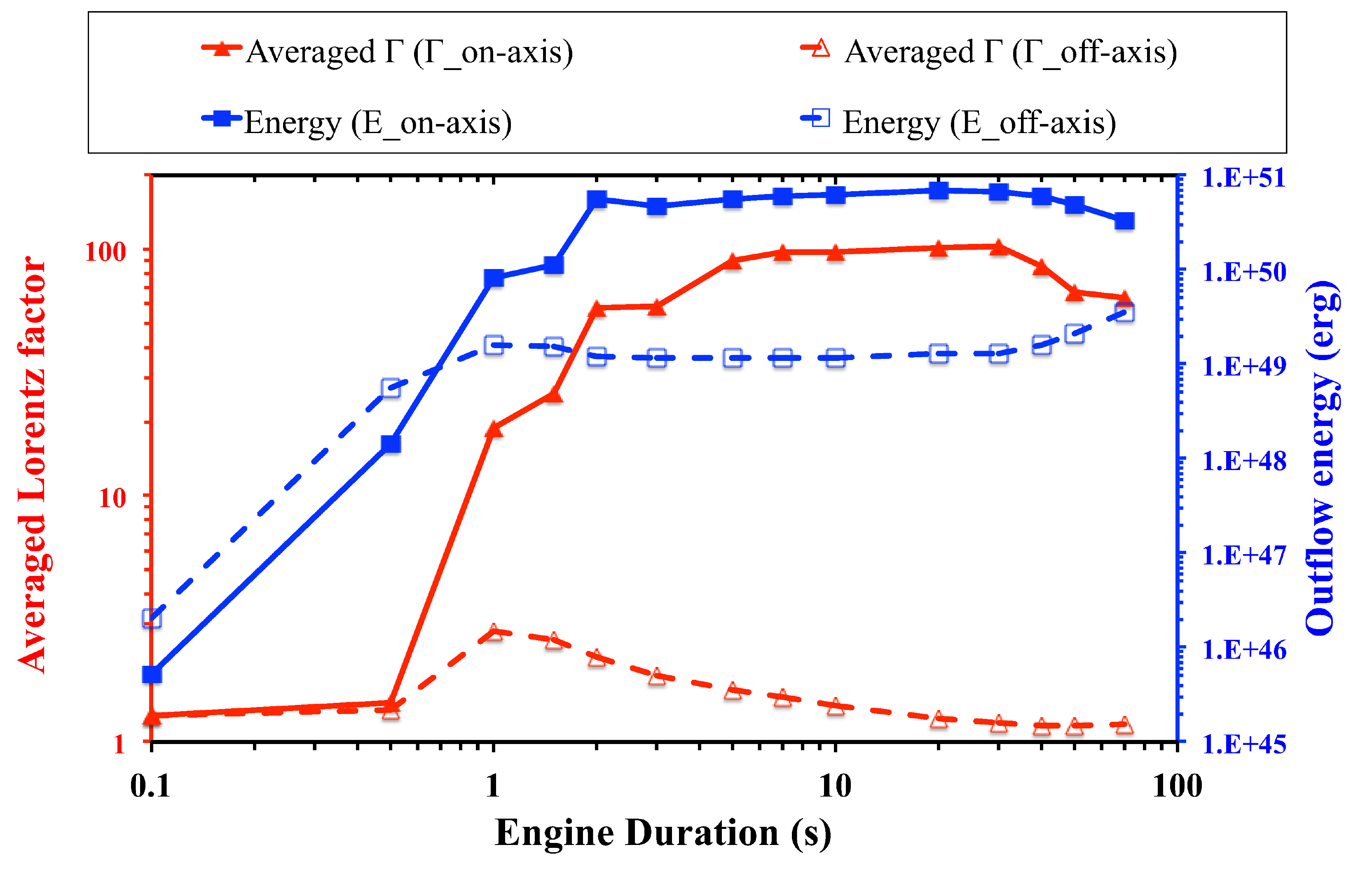} 
    %\vspace{4ex}
  \end{subfigure}%% 
  \begin{subfigure}%[b]{0.2\linewidth}
    \centering
    \includegraphics[width=0.4\linewidth]{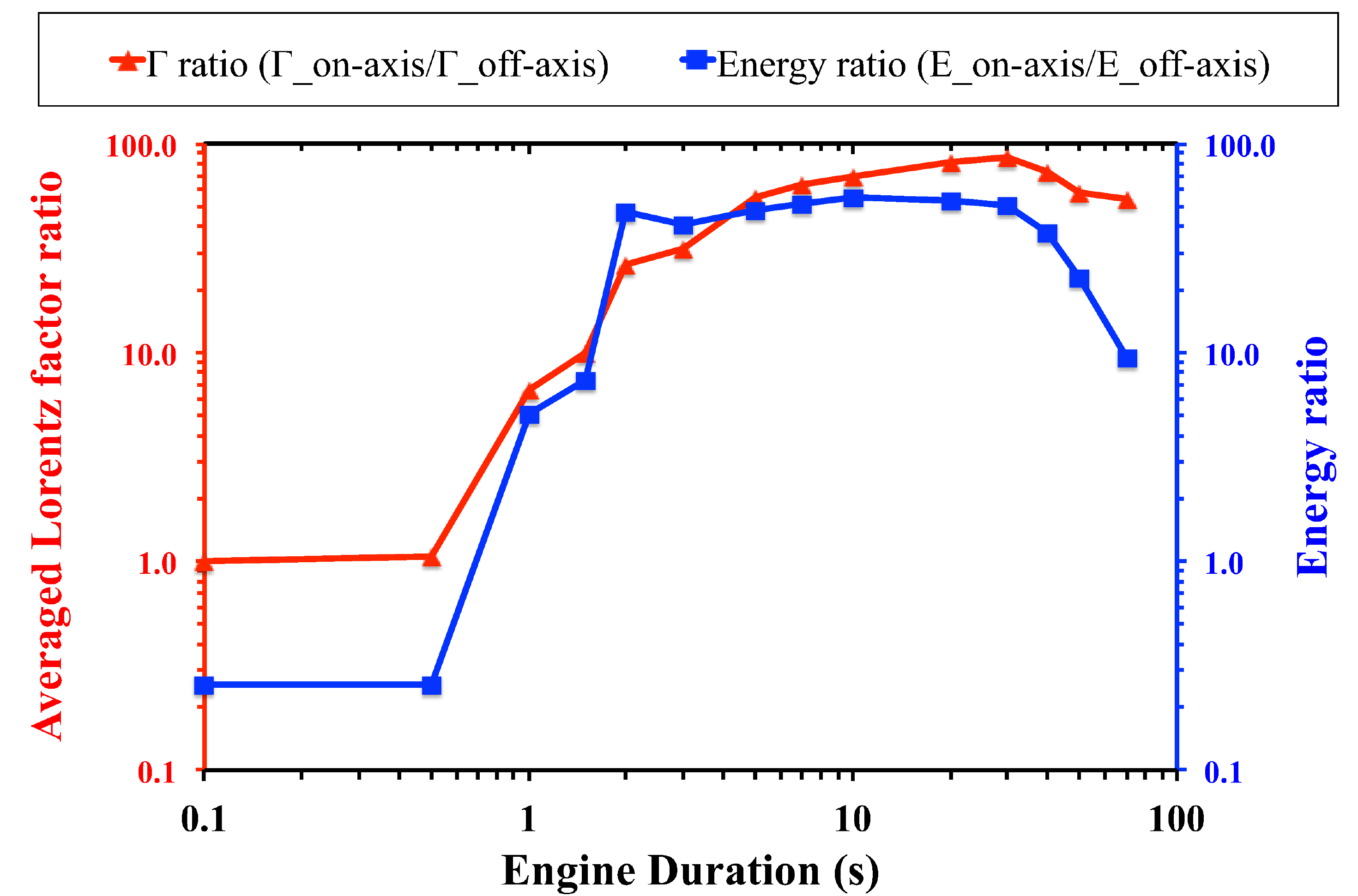} 
  \end{subfigure} 
   \vspace{4ex}
  \caption{On the left, the averaged Lorentz factor, in red, at the on-axis region ($0.125^{\circ}$) with filled triangles and solid line, and at the off-axis ($10^{\circ}$) with unfilled triangles and dashed line. In blue, the outflow total energy, at the on-axis ($0.125^{\circ}$) with filled squares and solid line, and at the off-axis region ($10^{\circ}$) with unfilled squares and dashed line. Both measured at $1.2\times10^{11}$ cm. On the right, the ratio of the averaged Lorentz factor at the on-axis over that at the off-axis, in red line and triangles. The blue line and squares show the ratio of the total energy at the on-axis over that at the off-axis.}
  \label{fig:collimation} 
\end{figure*}

Results are shown in figure~\ref{fig:collimation}. The figure's left panel gives more insights on how the relativistic and energetic properties differ for the different engine models. For brief engines, on-axis and off-axis quantities are of the same order. We can deduce that the outflow has, indeed, a quasi-isotopic angular distribution. In a jet context, collimation is very poor for brief engine models. With longer engines, longer than the breakout time, the gap between on-axis and off-axis quantities increase dramatically, with on-axis (energetic and relativistic quantities) about one order of magnitude higher than the same quantity in the off-axis region. This gap continues for all short and increases higher for intermediate engines, indicating that it is for these engine durations where jet collimation is the best. For long engines, the tendency of this gap is inversed, as it gets smaller. Still, it's higher than 10, indicating that the produced outflow is reasonably collimated. The general trend we found is that, the longer the engine duration the more relativistic and energetic the outflow is at the on-axis region, and the less relativistic and energetic the outflow is in the off-axis region; it seems simple and logical. However, what is surprising and new is that on the duration domain higher limit at long engines > 40 s, the tendency is slightly inversed. 

From the right panel in figure~\ref{fig:collimation}, the behaviour at our engine extreme limits (of duration), brief and long, implies a ``sweet spot'' in the middle; where the ratios, and thus the collimation is the best; which must be optimal for producing an energetic GRB's relativistic jet. Intermediate, followed by short engines make the finest jets in term of collimation. Long engines come next, making less collimation (but still successful jets). While brief engines come last, producing very poorly collimated jets, although the ratio, thus the collimation, improves as the engine duration increases. For the engine luminosities and durations presented here, the ``sweet spot'' is at $T_{inj}$ $\sim$ 5 -- 30 s, corresponding to engine luminosity per jet in the range: $1.6 - 10\times10^{50}$ erg s$^{-1}$ (a domain where both ratios are maximal). 

The poor collimation of jets from brief engines can be related to the quasi-isotropic properties of the unique jet phase launched by these engines, the precursor phase. The excellent collimation of jets from short and intermediate engines can be related to one main factor: The properties of the shocked phase, which dominates these two groups of engines, and which is the narrowest among the three phases (due to recollimation shocks). Finally, the inversed trend in long engines' jets might be related to the effective launch and large contribution of the third unshocked phase, which is characterized by gradually wider angular distribution, wider than the shocked phase (for more details on the unshocked phase for a 50 s engine see \citealt{2007RSPTA.365.1141L}). As jets are collimated by pressure from hot material in the stellar envelope, and as density of this surrounding material gradually decreases with time, the inversed trend of collimation (in unshocked phase and thus in long engines jets) makes sense.

This ``sweet spot'' may depend on progenitor properties, such as size, density profile, and the considered engine total energy. Our choice of the off-axis region here was at $10^{\circ}$. The choice of this off-axis region may influence the ratios, but we estimate that it will not affect the domain of the sweet spot and certainly not the general trend revealed here.

%%%%%
\subsection{Outflow acceleration efficiency}
\label{subsec:Outflow acceleration efficiency}

In this section we show the effect of the engine duration on the relativistic nature of the outflow's energy. Our different engine models have different tendencies (or efficiencies) at transmitting the engine energy into sub-relativistically or relativistically moving outflow. This efficiency depends on how the jet is able to form and expand. It also depends on the degree of baryon loading which reduce Lorentz factor. Some factors are suspected to play a potential role, parameters such as: timescale of the jet-star interaction, breakout time, and engine luminosity. In our engine models, the factors above differ as the engine duration varies from an engine to another. By comparing the different engine models, we show how the engine duration affects this efficiency, shaping the relativistic nature of the expanding outflow. 

We calculate fractions of the total engine energy that ends up in outflow expanding in the following relativistic domains: 1) Sub-relativistic: Fraction of the engine energy in material moving with a Lorentz factor $\Gamma$ > 1.005 (more 10\% the speed of light); 2) relativistic fraction: Fraction of the energy in material with $\Gamma$ > 10; and 3) highly-relativistic fraction: Fraction of the energy in material with $\Gamma$ > 100.

\begin{figure}
 \includegraphics[width=\columnwidth]{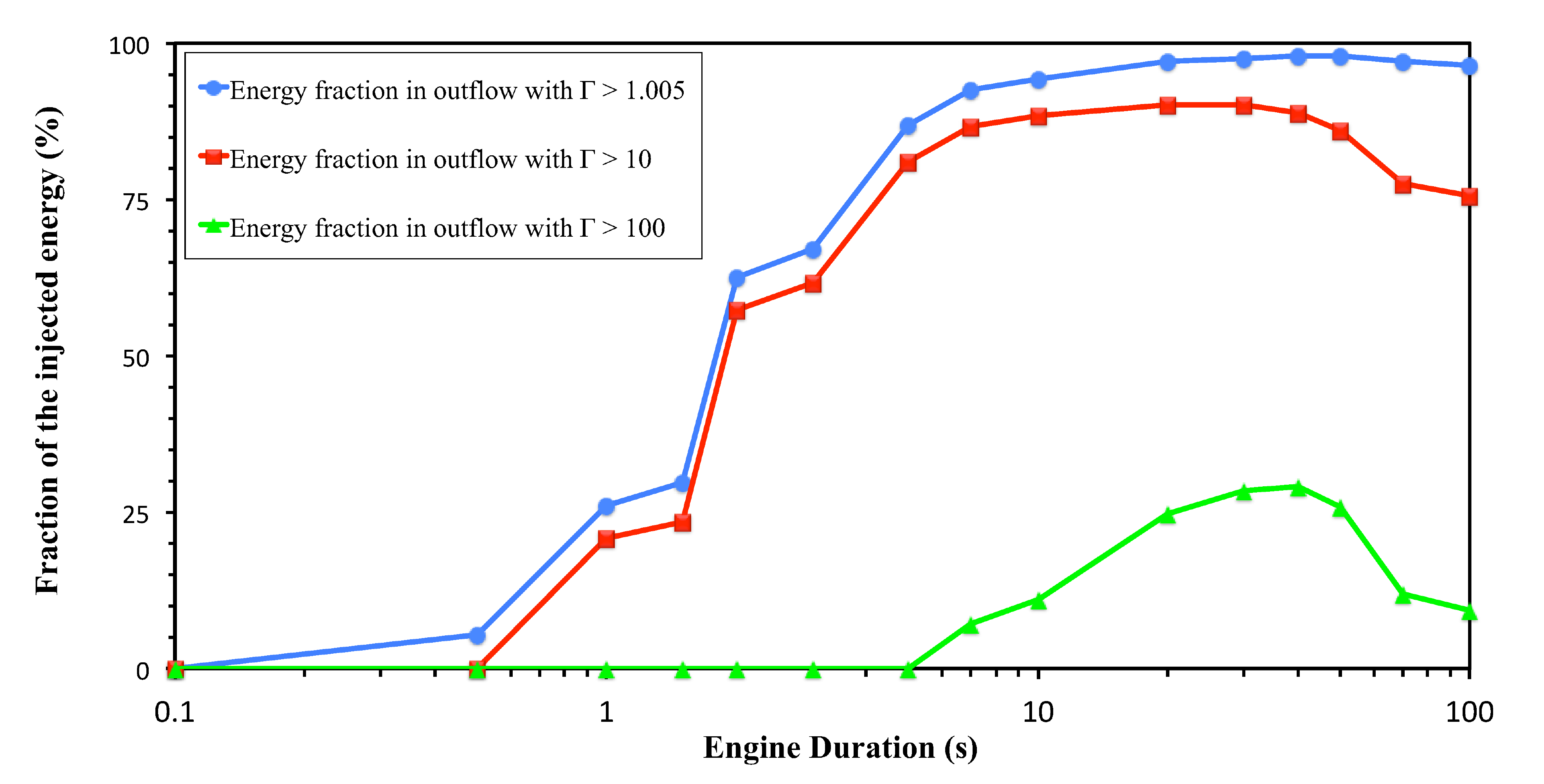}
 \caption{Fraction of total injected energy ($10^{52}$ erg) as found in different relativistically expanding outflows, as a function of the engine duration. In blue circles, total injected energy fraction in material with $\Gamma$ > 1.005, in red squares, energy fraction in material with $\Gamma$ > 10, and in green triangles, energy fraction in material with $\Gamma$ > 100. All measured at a radius of $1.2\times10^{11}$ cm.}
 \label{fig:acceleration}
\end{figure}

Figure~\ref{fig:acceleration} shows the effect of the engine duration on the relativistic nature of energy measured in the CSM. For brief engines, most of the engine energy is lost in poorly accelerated outflow, moving the slowest among the engine models presented here. Outflow expands with non-relativistic speeds ($\Gamma$ < 1.005). For longer brief engines, B010 and B015, acceleration is higher, but with a low efficiency; barely 25\% of the injected energy is successfully carried by outflow moving in the relativistic domain ($\Gamma$ > 10). For short engines, the efficiency is significantly higher, increasing with duration; from $\sim$60\% up to $\sim$85\% of engine energy is carried by material with $\Gamma$ > 10. The efficiency's increasing tendency with longer engine durations continues, up to intermediate engines domain, where the efficiency is the highest in all the three domains ($\Gamma$ > 1.005, $\Gamma$ > 10 and $\Gamma$ > 100), and roughly constant through intermediate engines' duration interval. For long engines, the tendency is slightly inversed, with less efficiency; still long engines remain capable of transmitting more than half of the engine energy into relativistically expanding outflow.

Thus, the efficiency at which engine energy is transferred to relativistic outflow, at engines' duration domain two extremities (brief and long), shows a ``sweet spot'', with increasing efficiency first, up to intermediate engines' domain where the efficiency is the highest, and roughly constant, before decreasing at long engines' duration domain. 

To understand the origin of this sweet spot behavior, we need to understand the reasons that led to lower efficiency at the two limits. At short durations limit, brief engines turn off before the breakout, resulting in a hot cocoon, instead of a relativistic jet. The cocoon largely mixes with the stellar envelope, which reduces its terminal Lorentz factor. At long durations limit, long engines (50 - 100 s) deliver the same total energy ($10^{52}$ erg). We suspect contamination of the baryon poor shocked jet ($\Gamma \gg 1$) by the baryon-rich stellar envelope ($\Gamma$ = 1) to lower the efficiency. Such contamination is favored by two main factors: 1) Lower jet luminosity leading to lower ram pressure, mixing, and then baryon contamination. 2) The long timescale during which the jet is in contact and in interaction with the stellar envelope, making it more subject to baryon contamination. Therefore, jets from intermediate engines with their quick breakout, high luminosity, and not very long timescale, are in an optimal domain that disfavor baryon contamination, and favor acceleration.

An interesting contrast here is that, for brief engines the relativistic nature of the produced jet outflow leaves most of the energy to power a non-relativistic event, possibly a SN explosion, with an accompanying soft GRB (\textit{ll}GRB). While for the other longer engines, most of the energy would be accelerated to contribute in producing a much powerful GRB, leaving much less energy in the non-realistic domain. This might explain why the best-recorded SN connections are Hypernovae accompanied by \textit{ll}GRBs, while typical GRBs are generally SN-less (Hamidani et al. 2017 in preparation). The behavior of long engines showing lower efficiencies is also interesting. This new finding questions the idea that very long collapsar engines are behind typical GRBs (supposing a total engine energy of $\sim10^{52}$ erg). This result is consistent with \citet{2013MNRAS.436.1867L} \& \citet{2012ApJ...749..110B} results, that used observational inputs to find that the contribution of long engines in observed GRBs samples (BATSE, Swift, etc.) is small.

%%%%%%%%%%%

\section{Conclusion}
\label{sec:Conclusion}

Using a 2D hydrodynamical relativistic code, we performed numerical simulations of a relativistic jet as powered by an accreting BH according to the collapsar model. It is the first time that such a wide engine duration domain was explored, which we separated into four groups (brief, short, intermediate and long). Our results allowed understanding more about the engine duration, a parameter not deeply studied so far both numerically and theoretically. We confirmed that engine duration could explain the large diversity of GRBs (as in \citealt{2012ApJ...750...68L}). The engine duration was found to dramatically affect almost all the jet properties. Also, we explored the link of engine duration to angular, temporal, and relativistic properties of expanding outflow. We confirm most of \citet{2012ApJ...750...68L} findings and present some interesting trends for the first time:

\begin{enumerate}

    \item Confined phase \& breakout time: Jet evolution inside the progenitor is different for the different engine durations. The general trend found here is that, in duration domain's two limits (brief and long), jets were slow at progressing and breaking out; while intermediate engines (short and intermediate) produced faster jets and shorter breakout times. The breakout time is a very important parameter that affects post-breakout jets and the resulting GRBs \citep{2002NewA....7..197R, 2011ApJ...739L..55B, 2011ApJ...740..100B}. These different breakout times produced other hydrodynamical differences and diversity in the emerging jets. Note that this is not the first work to show that very short duration jets have longer breakout times than intermediate durations, or that lower luminosity engines have longer breakout times. For instance, \citet{2012ApJ...750...68L} came to similar findings on breakout times (see Fig. 4). What is new here is the much wider duration domain, extending from very short to long durations.

    \item Jet phases \& light curve: We showed how different engine durations launch different jets, composed of different proportions of the three fundamental hydrodynamical phases: a) Brief engines launch jets composed of only the precursor phase, showing a ``cocoon-like'' properties. b) Short \& c) intermediate engines launch jets largely composed of the shocked phase with a short precursor phase; the shocked phase rich in shocks provides light curves rich in variability. d) Long engines launch jets composed of the three phases, with the smooth unshocked phase at its end; thus jet light curves; being a combination of all the three phases; were less variable. As GRB's prompt emission (and its light curve) is supposed to trance the evolution of the highly relativistic jet, GRBs' highly variable feature can be explained, most favorably in intermediate engines' duration domain. For one stellar model we could infer on the duration domain where jets are most efficient, durations seemingly consistent with observations and previous works. However, the discussion is obviously not complete to explain the cause of duration distribution in nature. As the engine physics is complex, the mechanism for a stable engine activity (or effective efficiency) may imply it to work for certain timescales. Parameters such as, core mass, angular momentum, magnetic field, viscosity, etc. strongly affect engine behavior (MW99 \& \citealt*{2008MNRAS.388.1729K}). For instance, as follows is how engines may shut-down (and thus duration distribution in nature is shaped): a) due to instabilities growing with time reducing jet propagation velocity (\citealt*{2016MNRAS.456.1739B}), b) following the spin-down time in a magnetar engine (\citealt*{2001ApJ...552L..35Z}), c) small viscosity leading to a rapid decline in jet luminosity (\citealt{2008MNRAS.388.1729K}), etc.

    \item Angular distribution \& collimation: Our study on angular distribution for different durations is the first. We showed that brief engines produce poorly collimated outflow, in the form of a quasi-isotropically expanding material. A general trend was that, for longer engine duration collimation increase significantly, and the produced jet's outflow and energy were much narrowly distributed in the on-axis region (ideal for a GRB-jet). However, for long engines this trend was slightly inversed, although the jet remained reasonably collimated. Thus, the result is a ``sweet spot'' for engine durations were the collimation is the best, in the domain of short and intermediate engines: $\sim$ 5 -- 30 s. 

    \item Lorentz factor \& Energy fraction carried by relativistic outflow: Again, we showed that there are engine durations for which the launch of GRB's extremely relativistic and energetic jets are favored. Brief engines were found incapable of efficiently accelerating the outflow to relativistic velocities necessary to explain GRBs. Instead most of the engine energy was lost in non-relativistically expanding outflow (< 10\% the speed of light). With longer engine durations, the efficiency at transmitting engine energy to highly relativistic speeds increased, and reached its maximum for intermediate engines, in a ``sweet spot'' ($\sim$ 10 -- 30 s), making the most relativistic and energetic jets of our sample. For long engines, the trend was inversed and the efficiency started to drop, although the jet remains considerably accelerated and energetic for our longest engine. 

\end{enumerate}

Considering the above results (i,ii, iii \& iv), we can reach an important conclusion on GRBs. For a typical collapsar; in a typical 25$M_{\sun}$ Wolf-Rayet star, with a typical collapsar explosion energy of $\sim10^{52}$ erg; to reproduce a standard GRB with its energetic, relativistic, beamed, and variable features, there is a favorable engine duration domain. We found that this duration domain is around intermediate engines $\sim$10 -- 30 s; as the following favourable conditions are met: 1) relativistic breakout, and $T_{inj} \gg T_{breakout}$ necessary for successful jets (an essential requirement, as previously argued in \citealt{2012ApJ...749..110B}); 2) jet temporal evolution rich of variability, capable of reproducing GRBs' high variability; 3) best collimation (sweet spot); and 4) efficiently energetic and relativistic outflow (sweet spot). These features come from the shocked phase, which is the dominating phase for intermediate engines.

In the short duration limit (brief engines) where $T_{inj} < T_{breakout}$, the product would rather be a non-relativistic event (core-collapse SN; as found in \citealt{2012ApJ...750...68L}) or a soft GRB (\textit{ll}GRB) likely to be associated with a powerful SN (Hamidani et al. 2017 in preparation). In the limit of long engines ($\sim$ 50 - 100 s), the production of GRB event might still be possible, although less likely than in the domain of intermediate engines. However, as the quality of collimation and relativistic acceleration keep decreasing beyond intermediate engines, we argue that very long engines (> 100 s) are not expected to be efficient enough to power GRB. Hence, regardless of theoretical existence of very long accreting models, very long / ultra-long engine models might present hydrodynamical problems at explaining GRBs (unless larger engine energy and thus luminosity, or bigger progenitors, available). Note that our conclusions were drawn considering a collapsar engine typical energy budget of $10^{52}$ erg, and that larger energy budget might affect the results, but not the general trend. This trend (where a lower duration limit and an upper duration limit exist for GRB engines) might gives clues to explain the duration distribution of LGRBs (its domain and why it peaks around a few tens of seconds), if more diverse stellar models are considered to account for diversity in the universe.

Our result, regarding the GRBs' ideal engine duration domain, is consistent with two previous works, each with a different approach and with observational inputs. First, \citet{2012ApJ...749..110B} estimation of engine duration distribution using an analytical formula of breakout times (and assuming that $T_{90} = T_{inj} - T_{breakout}$) which suggested that GRB engines longer than $\sim$ 100 s are very unlikely in nature. Second, \citet{2013MNRAS.436.1867L} which after assuming a constant SFR in the universe, $T_{90} = T_{inj} - T_{breakout}$, and inputs from both numerical simulations and BATSE observations, indicated that BATSE GRBs seem to come from engines duration domain around $\sim$10 -- 20, and mostly not long engines. Thus, our study is an additional support to the collapsar model and the scenario of a relativistic jet from a massive WR star. 

%%%%%%%%%%%

\section*{Acknowledgements}
% Entry for the table of contents, for this guide only
\addcontentsline{toc}{section}{Acknowledgements}

We gratefully thank Toshiazu Shigeyama, Kunihito Ioka, Norita Kawanaka \& Shigehiro Nagataki for the many helpful discussions. We also thank Tomonori Totani, Miguel Aloy, Akira Mizuta and Shingo Hirano for the valuable comments. We are also very grateful to Aaron C. Bell for improving the quality of this paper. This research was supported by the MEXT of Japan, for which we are gratefully thankful. Numerical computations were carried out on Cray XC30 at Center for Computational Astrophysics, National Astronomical Observatory of Japan. This work has been partly supported by Grants-in-Aid for Scientific Research (17K05380) from Japan Society for the Promotion of Science and Grant-in-Aid for Scientific Research on Innovative Areas (26104007) from the MEXT in Japan.

%%%%%%%%%%%%%%%%%%%%%%%%%%%%%%%%%%%%%%%%%%%%%%%%%%



\bibliographystyle{mnras}
\bibliography{mnras} % if your bibtex file is called ?example.bib

%\begin{thebibliography}{99}
%\bibitem[\protect\citeauthoryear{Author}{2013}]{author2013}
%\bibitem[\protect\citeauthoryear{Jones}{2015}]{jones2015}
%Jones C.~D., 2015, Journal of Interesting Stuff, 17, 198
%\bibitem[\protect\citeauthoryear{Smith}{2014}]{smith2014}
%Smith A.~B., 2014, The Example Journal, 12, 345 (Paper I)
%\end{thebibliography}

%%%%%%%%%%%%%%%%%%%%%%%%%%%%%%%%%%%%%%%%%%%%%%%%%%

%%%%%%%%%%%%%%%%% APPENDICES %%%%%%%%%%%%%%%%%%%%%

\appendix

\section{Code Testing}
\label{sec:Code Testing}

In order to estimate the accuracy of our numerical code, we carried out several calculation tests. Here we present a test of the code by solving Riemann problem. We calculated the propagation of a blast wave, generated by left and right phases initially detached by diaphragm, the so-called shock-tube problem. Since it is difficult to get analytical solution in the spherical case, we reproduced the 1D Riemann problem as in \citet{2002A&A...390.1177D} who proposed the computational convergence by comparing between a solution with coarse grid and with well-resolved fine grid.

Numerical grids are set as $0 \leq r \leq 1$, and the speed of light c is set to 1. All the physical quantities are dimensionless. We divided the numerical domain in r direction by $N_r = 200$ zones and by $N_r = 800$ zones for the ``coarse'' and ``fine'' grid cases, respectively. In $\theta$ direction, $N_{\theta} = 16$ zones ranging in $0^{\circ} < \theta < 90^{\circ}$ for both cases. The initial condition is given as follows:

\begin{eqnarray}
 (\rho,\upsilon_r,p)  = \begin{cases} (1,0,1000), & \mbox{For } r \leq 0.4 \\ (1,0,1), & \mbox{For }  r \geq 0.4 \end{cases}
\end{eqnarray}

\begin{figure}
 \includegraphics[width=\columnwidth]{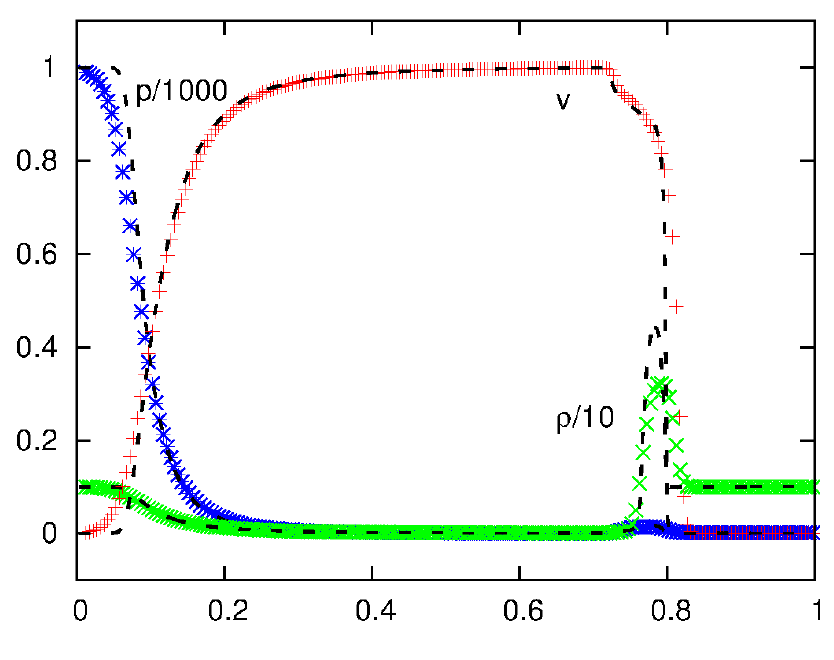}
 \caption{Numerical solutions of one-dimensional spherical Riemann problem corresponding to the physical quantities $\rho/10$, $\upsilon_r$, $p/1000$, as marked. Black dashed line is the well-resolved numerical solution where the computational domain is divided by $N_r = 800$ zones. Coarser solution, corresponding to a domain divided by $N_r = 200$ zones, is plotted with marks.}
 \label{fig:test1}
\end{figure}

Explosion generates outgoing and incoming shocks. Figure~\ref{fig:test1} shows the coincidence of those solutions. This proves the robustness of our numerical code for the propagation of a blast wave. For a comparison see \citet{2006ApJ...651..960M} Fig. 21.

\section{Effect of resolution}
\label{sec:Effect of resolution}

\begin{figure*}%[ht] 
    \vspace{4ex}
  \begin{subfigure}%[b]{0.2\linewidth}
    \centering
    \includegraphics[width=0.4\linewidth]{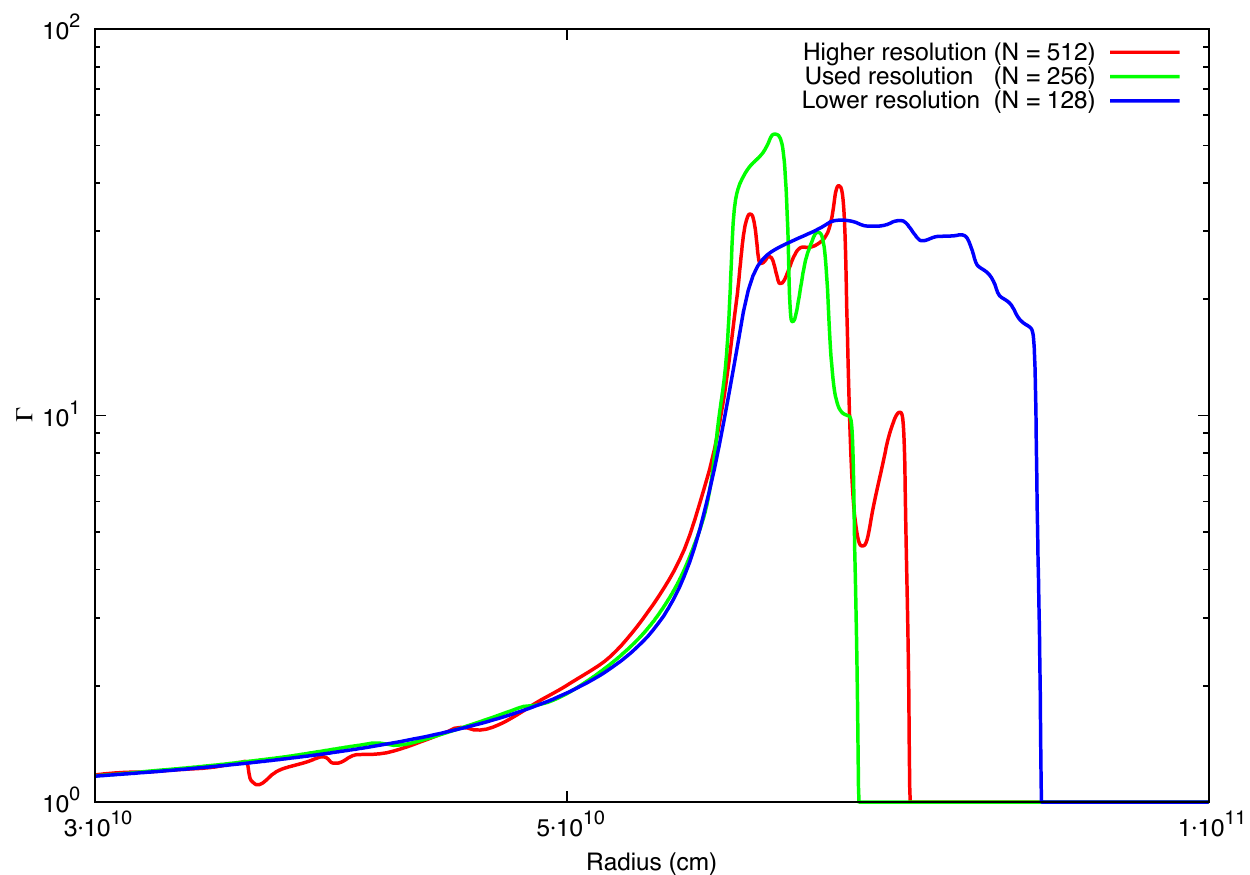} 
    %\vspace{4ex}
  \end{subfigure}%% 
  \begin{subfigure}%[b]{0.2\linewidth}
    \centering
    \includegraphics[width=0.4\linewidth]{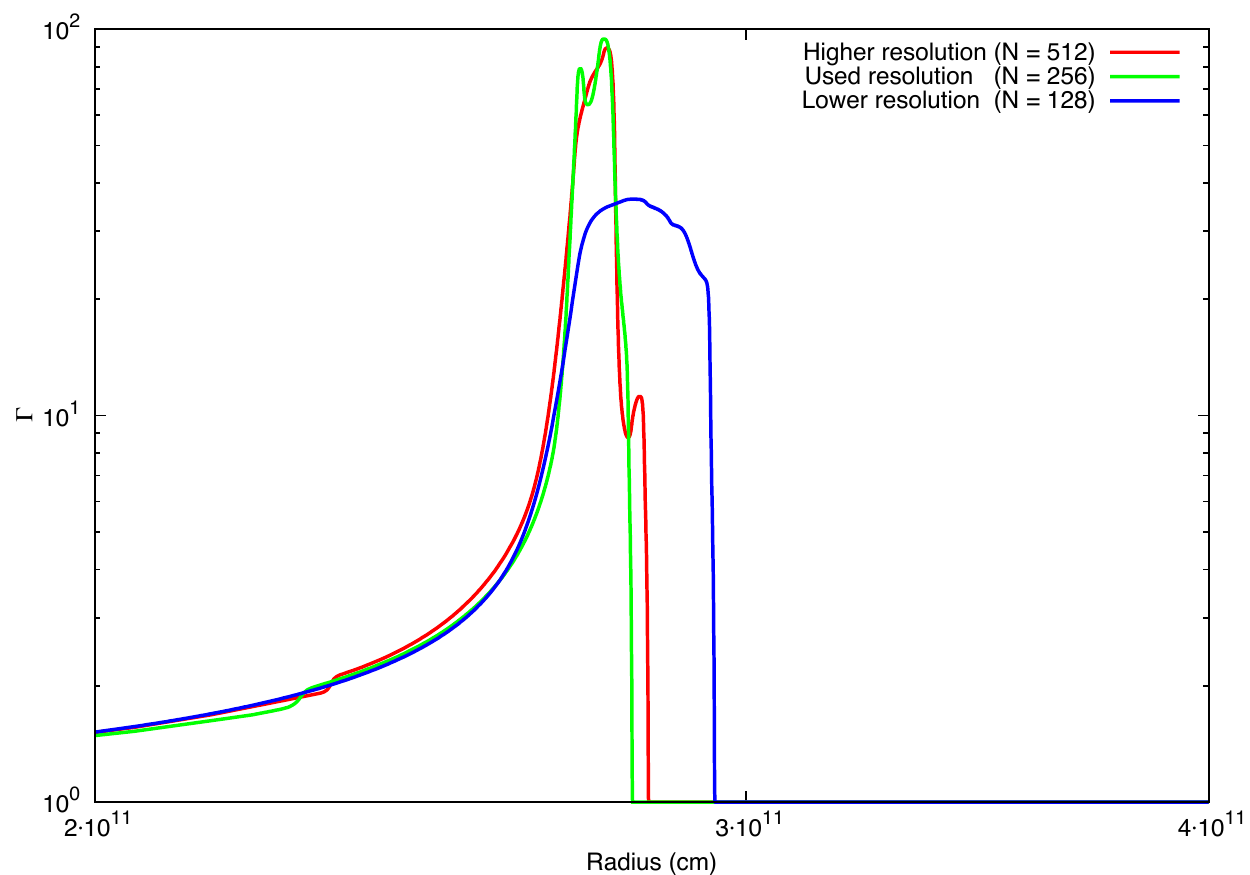} 
  \end{subfigure} 
   \vspace{4ex}
  \caption{Comparison of calculations with different angular resolutions, at different times (On the left at 3.0 s just after the breakout, on the right at 10.0 s, a few seconds after the breakout). ``Higher resolution'' calculation uses 512 grids in $\theta$ direction, the ``used resolution'' in the calculations of this study uses 256, and the ``lower resolution'' has an angular grid of 128. The used resolution (256) and the higher one (512) show similar profiles, and thus 256 is fairly good enough.}
  \label{fig:test2} 
\end{figure*}

We also tested the angular resolution of our simulations. Since, the jet has a collimated structure, it's needed to check whether angular grids can resolve the collimated narrow structure. The test is on whether the considered angular resolution in our study is good enough. We compared the same calculation in three different grid resolutions: ``higher resolution'' $N_{\theta} = 512$ angular meshes, the ``used resolution'' in our study $N_{\theta} = 256$, and ``lower resolution'' $N_{\theta} = 128$. The ``used resolution'' is comparable to, or finer than, some previous studies (e.g. \citealt{2009ApJ...699.1261M}). Other jet initial parameters are all the same, as in B010. Comparison of the three calculation is showed in figure~\ref{fig:test2}. There is a dramatic difference in the jet structure between the ``lower resolution'' and the two higher resolutions. The calculation with $N_{\theta} = 128$, seems to lead to some lose of fine structure in the jet, a structure that 256 and especially 512 are displaying. 512 and 256 resolutions are very similar; almost converging, suggesting that 256 grids are reasonably good enough, and the use of 512 is not strongly needed and would not significantly change the jet structure, and thus the results. 

\section{Comparaison with ML07}
\label{sec:Comparaison with ML07}

The calculation we present here is to compare the output from our calculation to that of ML07's main models, 16TIg5 and t10g5. ML07 used an engine duration of $T_{inj} = 50$ s, delivering a total energy of $E_{tot} = 5.32\times10^{52}$ erg, with the injection nozzle situated at $R_{in} = 10^9$ cm. We carried out a simulation with the same above: $R_{in}$, $T_{inj}$ and $E_{tot}$, as in ML07. The aim of this comparison is to test our code output and make sure that it does not have any deficiencies, and thus is as correct as the FLASH code used in ML07. Thus, we carried a simulation with identical engine properties to that of ML07's 16TIg5 and t10g5 models, with engine luminosity per jet, of $5.32\times10^{50}$ erg (table \ref{table:engines}). We have to mention that the simulation still presents some minor differences in comparison to ML07 and that the two simulations are not fully identical. Minor differences such us ratio of internal over rest mass energy, progenitor, EOS, resolution, etc. Nevertheless, that did not prevent our results from being in an excellent agreement with those of ML07. Table \ref{table:breakouts} last line represents the breakout times of the three phases, small differences in the breakout times exist more likely due to difference in the progenitor, but the difference is minor and the breakout times are in the same order (for a comparison see 16TIg5 model in ML07's table 2).

\begin{figure*}%[ht] 
    \vspace{4ex}
  \begin{subfigure}%[b]{0.2\linewidth}
    \centering
    \includegraphics[width=0.4\linewidth]{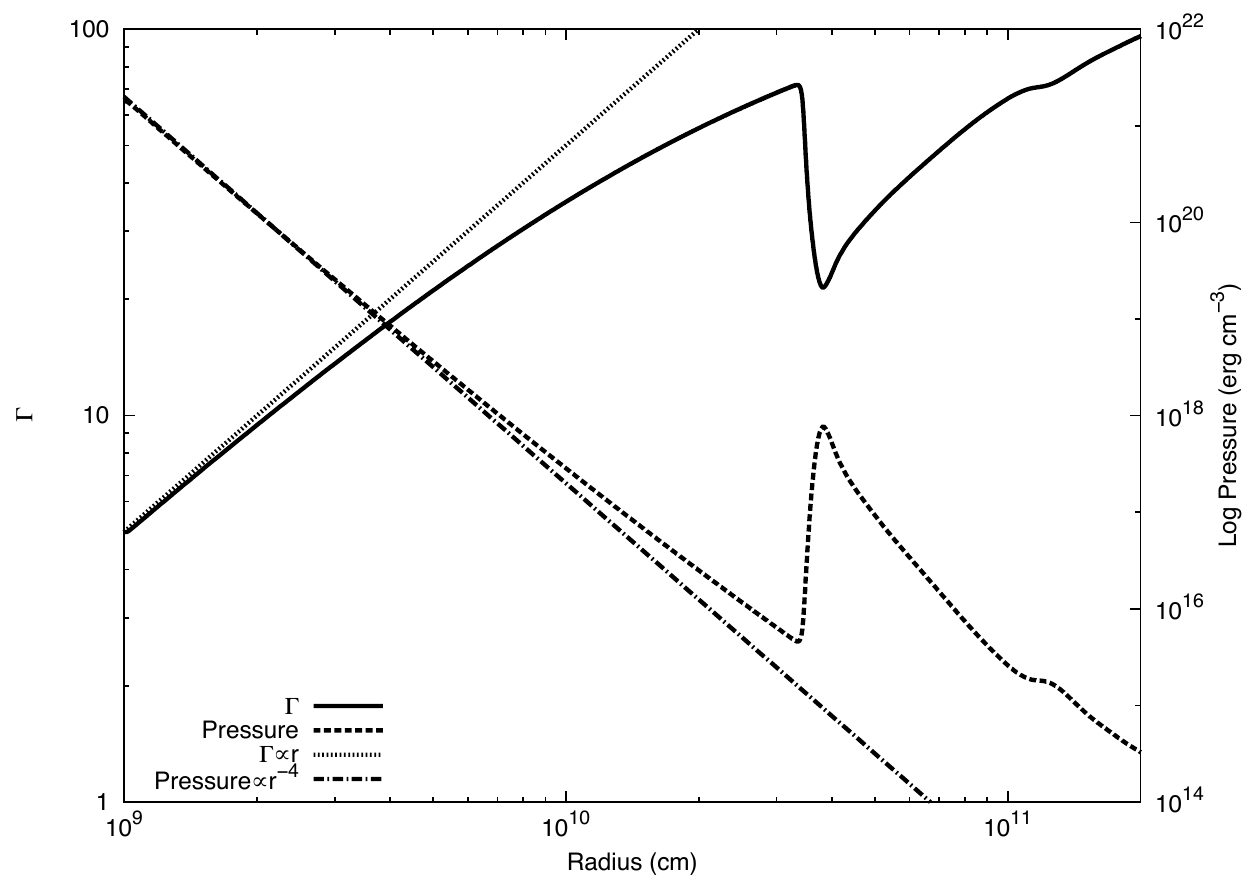} 
    %\vspace{4ex}
  \end{subfigure}%% 
  \begin{subfigure}%[b]{0.2\linewidth}
    \centering
    \includegraphics[width=0.4\linewidth]{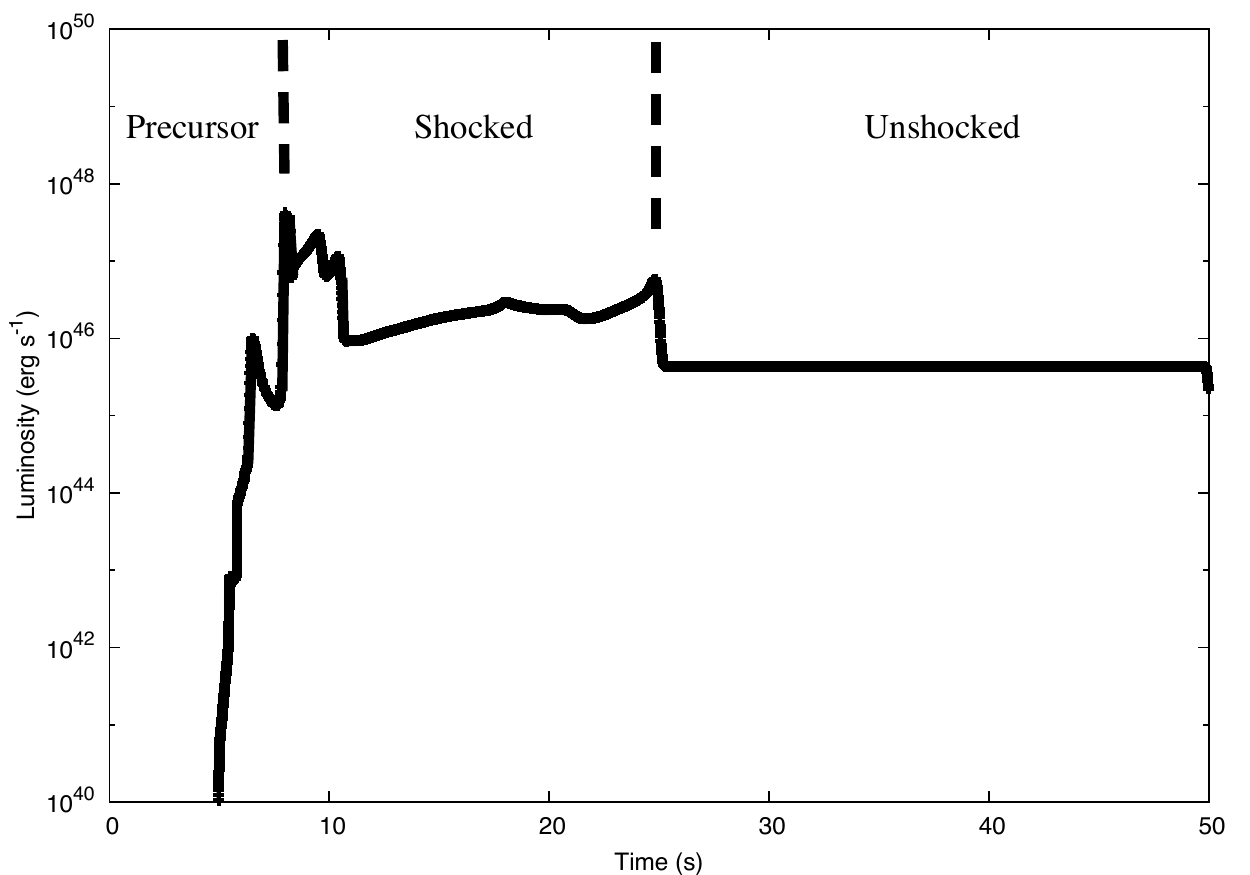} 
  \end{subfigure} 
   \vspace{4ex}
  \caption{On the left panel, Lorentz factor (left y-axis), and pressure (right y-axis) as a function of the radius, in solid line, and dashed line, respectively. Both quantities are computed at the jet-axis region, the innermost angular grid. The dotted line, and dotted dashed line, shows the theoretical prediction considering a free-streaming jet pressure-dominated jet, with Lorentz factor and pressure, respectively. This approximation holds very well as long as the jet outflow is not at highly relativistic speeds ($\Gamma \ll 100$). On the right, the energy flux over time along the jet on-axis. The energy was calculated at $1.2\times10^{11}$ cm, as in ML07. Dashed lines show the transition times between the three phases of the jet: precursor to shocked, and shocked to unshocked phase.}
  \label{fig:unshocked} 
\end{figure*}

Next in figure~\ref{fig:unshocked}, left panel, we present the properties of the unshocked material at the moment of the breakout, 25.0 s after the start of the calculation. The core of the unshocked jet is, indeed, in agreement with the theoretical prediction of a free-streaming jet (\citealt{2007RSPTA.365.1141L} and the references within). It is also in excellent agreement with ML07's 16TIg5 model results (see ML07 figure 10). With Lorentz factor proportional to the radius, and the pressure proportional to $r^{-4}$, the unshocked jet in our calculation is well in agreement with both theoretical predictions and the previous work presented in ML07 and \citet{2007RSPTA.365.1141L}. We have to mention that, as in ML07, with Lorentz factor getting closer to $\sim$100, derivation from the theoretical prediction increases, this is due to, as explained in ML07, the fact that at such highly relativistic speed, the approximation of the flow being pressure-dominated no longer holds. For a comparison of this figure with previous works see: \citet{2000ApJ...531L.119A} Fig. 2, \citet{2003ApJ...586..356Z} Figs. 4, 5, and 6, \citet{2009ApJ...699.1261M} Figs. 6 and 7, and Fig. 10 in ML07.

Figure~\ref{fig:unshocked} right panel shows the outflow energy flux along the jet axis over time for our calculation. The three phases are clearly identifiable. Here again, the temporal and energetic properties demonstrate that our calculation is in excellent agreement with that of ML07, despite the few minor differences in the calculation setting. Considering the isotropic equivalent luminosity would add a factor of $\sim8\times10^5$ bringing the energy to the same order as in ML07 (for a comparison, see ML07 figure 4).

\begin{figure*}%[ht] 
    \vspace{4ex}
  \begin{subfigure}%[b]{0.2\linewidth}
    \centering
    \includegraphics[width=0.4\linewidth]{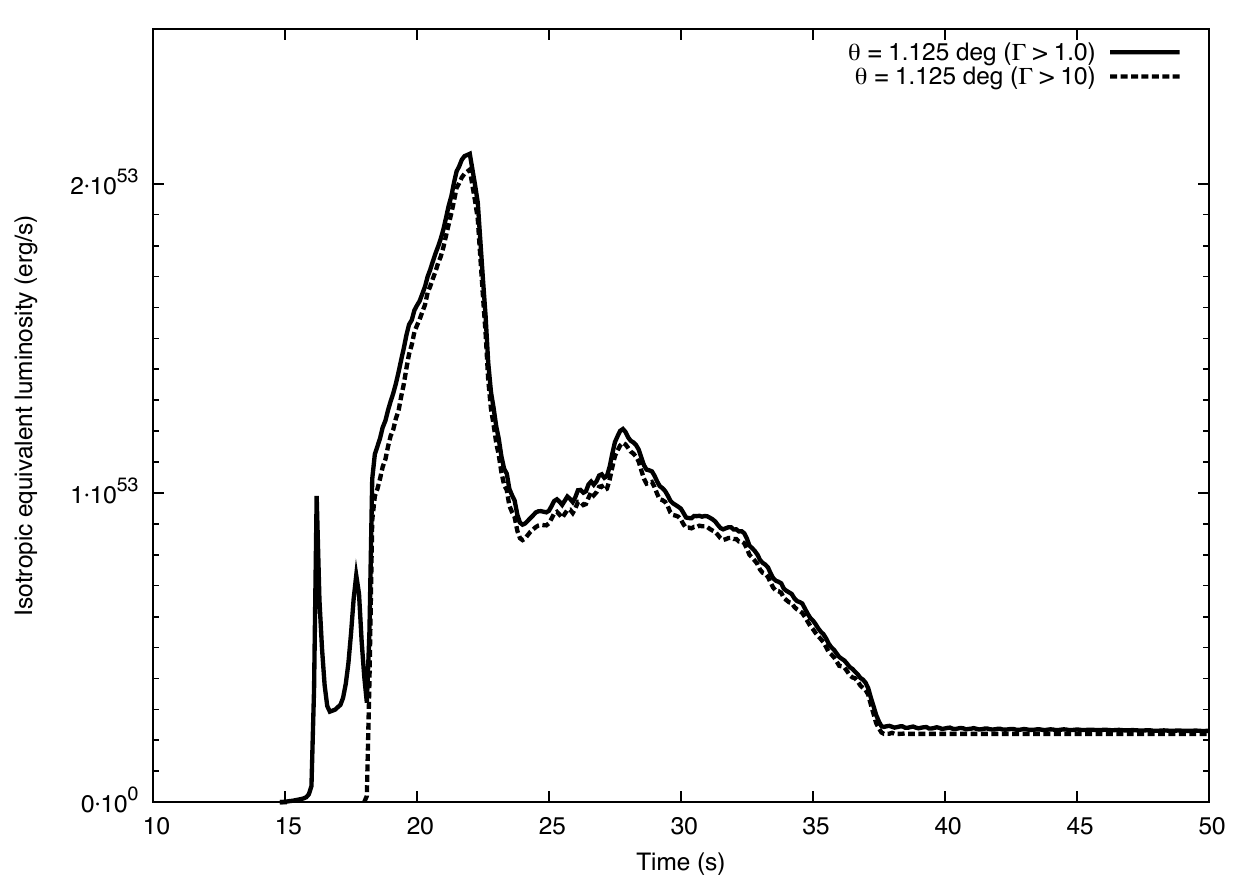} 
    %\vspace{4ex}
  \end{subfigure}%% 
  \begin{subfigure}%[b]{0.2\linewidth}
    \centering
    \includegraphics[width=0.4\linewidth]{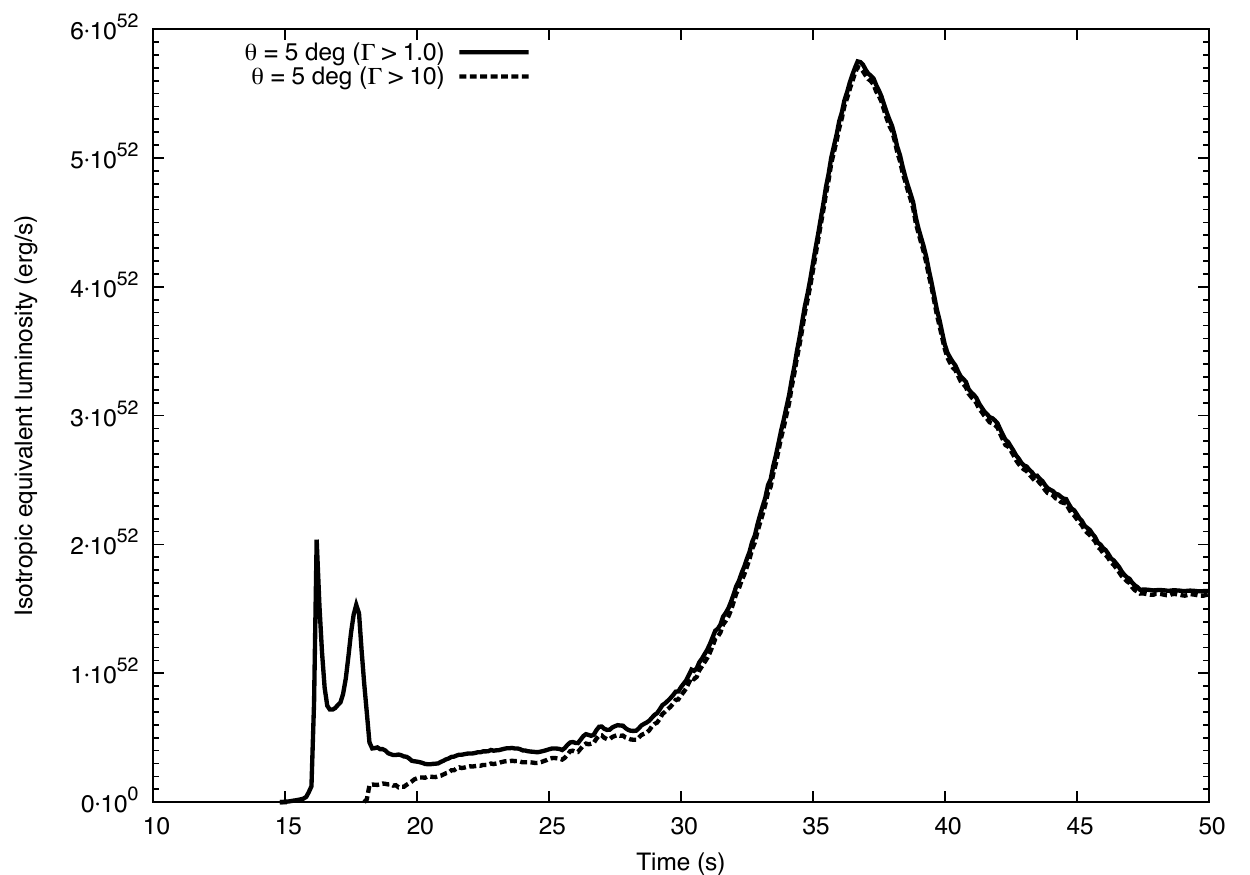} 
    %\vspace{4ex}
  \end{subfigure} 
  \begin{subfigure}%[b]{0.2\linewidth}
    \centering
    \includegraphics[width=0.4\linewidth]{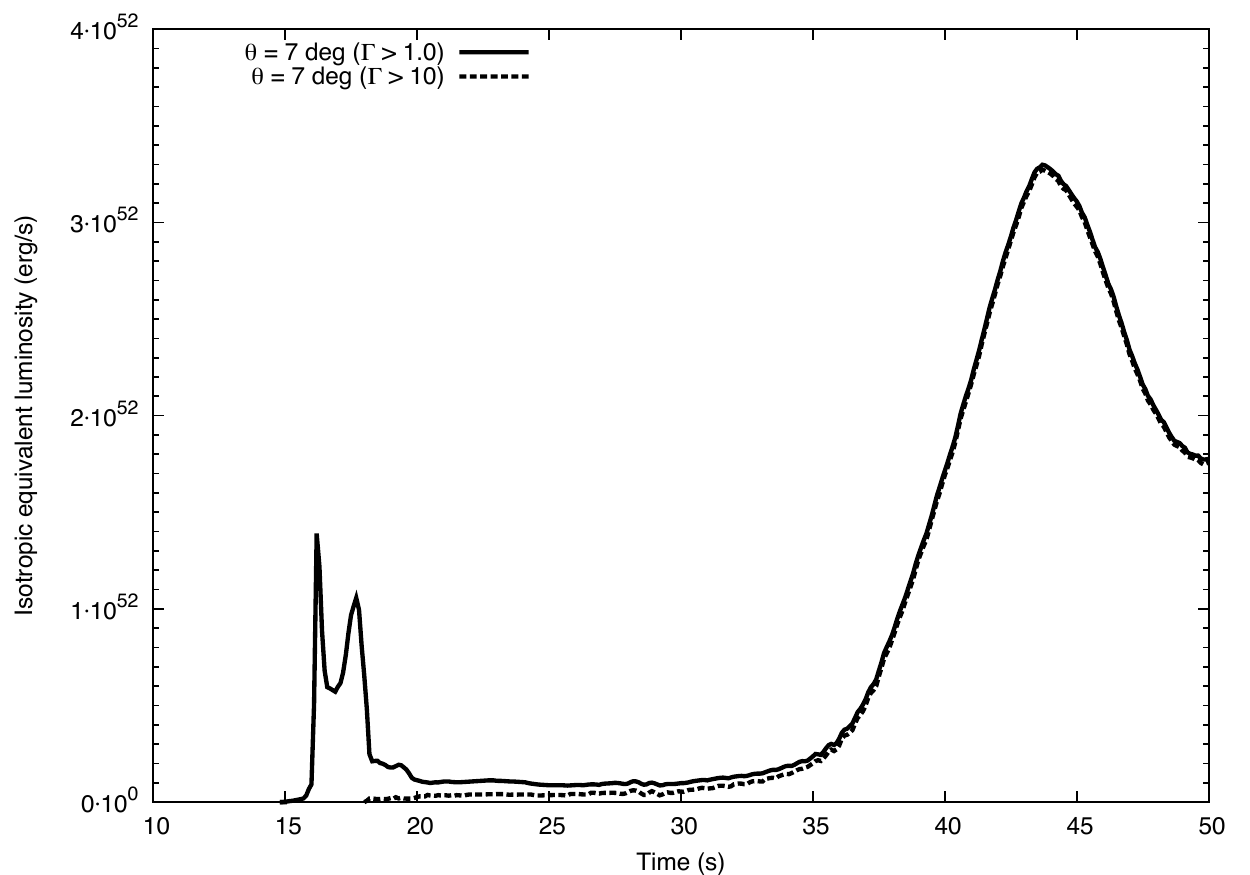} 
  \end{subfigure}%%
  \begin{subfigure}%[b]{0.2\linewidth}
    \centering
    \includegraphics[width=0.4\linewidth]{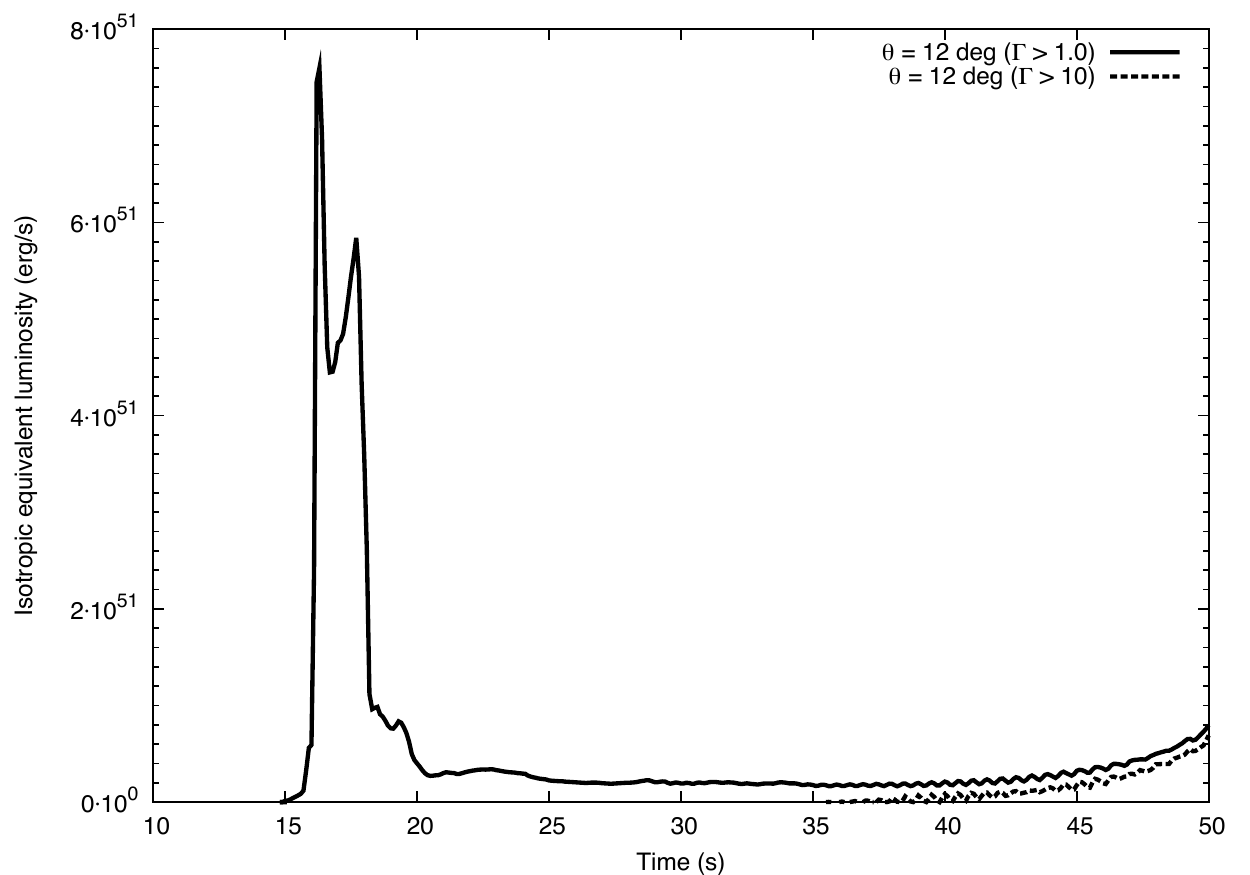} 
  \end{subfigure} 
  \caption{Light curves for a similar engine to that of 16TIg5 and t10g5 in ML07. As in ML07's figure 12, energy flux over time is shown for four different viewing angles. Solid and dashed lines are for material with a minimum Lorentz factor of 1.0 and 10, respectively. The four light curves are plotted at angles: $1.125^{\circ}$ (top left), $5^{\circ}$ (top right), $7^{\circ}$ (bottom left), and $12^{\circ}$ (bottom right). These light curves were estimated as explained in \S \ref{sec:Data processing procedure}, and as in ML07 (figure 12 for a comparison).}
  \label{fig:ML07-12} 
\end{figure*}

Finally, figure~\ref{fig:ML07-12}, presents light curves calculated as explained in \S \ref{sec:Data processing procedure}, following ML07 method. Apart from the light curve at $1.125^{\circ}$ showing minor difference although it remain very similar, the light curves are identical to those presented in ML07 for t10g5 model (see ML07's figure 12 for a comparison). The slight difference in light curves at $1.125^{\circ}$ is most likely due to ML07's considerably higher angular resolution at that region, near the jet axis. One other difference is in the precursor's Lorentz factor, which is lower than 10 in our calculation, at the difference of that of ML07's t10g5. This is most likely due to difference in the progenitor, which is expected to strongly affect this phase, rather than difference in the numerical method or in the physics. In our case the stellar model is realistic \citep{2005ApJ...633L..17U}, where in ML07's t10g5 it is a power-law stellar model. Nevertheless, apart from these two minor differences, the energy range, dead times, temporal properties, are at excellent agreement, allowing us to conclude that our code is robust and as safe as that of ML07, and confirming ML07 results.

From the above tests and calculations, we conclude that our simulations do not suffer from any numerical problems. We can also confirm that our simulation setting, such as the choice of resolution, doesn't miss the jet structure, and thus is appropriate. Also, the comparison with ML07's 16TIg5 and t10g5 models shows that our code's output and numerical treatment is at excellent agreement with that of ML07. Thus we can conclude that our numerical code is robust enough and fully appropriate for the kind of study presented here.

%%%%%%%%%%%%%%%%%%%%%%%%%%%%%%%%%%%%%%%%%%%%%%%%%%

% Don't change these lines
\bsp	% typesetting comment
\label{lastpage}
\end{document}